\documentclass[amsmath,amssymb, aps,prl,longbibliography ]{revtex4-2}
\usepackage{graphicx}
\usepackage{dcolumn}
\usepackage{bm}
\usepackage{caption}
\usepackage{subcaption}

\usepackage{amssymb}
\usepackage{caption}
\usepackage{subcaption}
\usepackage{amsmath}
\usepackage{xcolor}
\usepackage{lipsum}
\usepackage{graphicx}
 \usepackage[normalem]{ulem} 

\begin{document}

  \title{\textbf{Enhanced quantum transport in bilayer two-dimensional
      materials}}

\author{Jos\'e Campos-Mart{\'i}nez }
\email{jcm@iff.csic.es}
\author{Marta I. Hern\'andez}

\affiliation{%
 Instituto de F{\'i}sica Fundamental \\
 Consejo Superior de Investigaciones Cient{\'i}ficas \\
 Serrano 123, E-28006 Madrid, SPAIN}%

\begin{abstract}
   Two-dimensional (2D) materials have been proposed, among many
  other applications, as an efficient tool for
  the separation of atomic and molecular species and their
  corresponding isotopes, given the confinement provided
  by their subnanometric dimensions.  In this work we present 
  three dimensional quantum  wave
  packet calculations revealing an enhancement in the quantum
  transport in bilayer  over monolayer graphdiyne membranes,
  one of the most popular 2D materials which is commonly employed for
  this purpose.
  Besides, resonances emerge superimposed over the
  typical monolayer profile for transmission probabilities, a
  feature that
  is general to other bilayer nanoporous 2D heterostructures and
  that shows a strong dependence on the interlayer separation.
\end{abstract}

\maketitle



\section{Introduction}
The research devoted to two-dimensional (2D) materials has grown
exponentially
\cite{research-prog-2d-mat-2025,ares_recent_2022,mat-23-syn-chem-17}
spurred by the great amount of applications that are foreseen
\cite{gr2-rev-fundamental-and-applied,zheng_two-dimensional_2023},
along with the
continuous development of new and more sophisticated ways of synthesizing
these nanostructures\cite{kang_graphyne_2019,li_graphynes_2023,zhou_gas_2022,gr2-review-22}.
Within these new 2D  materials, graphynes
\cite{gr2-review-22,review-gr2-and-applicat-23,gr-rev-syn-prop-app-23-flatchem} 
are very interesting nanoporous carbon layers, with a long research tradition.
In fact graphynes were firstly theoretically
predicted in 1987\cite{baughman_structureproperty_gr-1987} and later
synthesized in 2010\cite{li_architecture_2010}.  Since then, the
experimental procedures have been continuously
improving\cite{xu_transparent_2019},
and as today, we can genuinely say that these 2D materials can
be routinely prepared to meet almost any desired application
\cite{yasuda_efficient_2022,pan_graphdiyne_2021,li_graphynes_2023}
and, even if there are still issues concerning low yields and costly
operations \cite{pgrogress-gr2-gas-separa-25}, there are also indications
that scalable synthesis might be on the
horizon to possibly take these materials from the lab to the industry level
\cite{synthe-graphy-22-classifi,gr2-multlayer-water-selec-25}.
From a fundamental point of view graphynes can
  be described as materials containing $sp^1$ hybridized linear chains,
  as linear polyyne chains $-(C\equiv C)n-$, and
  they can be classified as\cite{synthe-graphy-22-classifi} graphynes-$n$
  depending on the number, $n$, of acetylenic bonds and $x, y, z$ graphynes
  with at least some nonaromatic $sp^2$ carbons, although the names
  $\alpha-, \beta-,\gamma-$ graphynes have been also
  employed\cite{li_graphynes_2023} to distinguish between
  different families.    
Among them, one of the 2D layers more largely studied is known as
graphdiyne (GDY)\cite{gr2-rev-fundamental-and-applied,li_graphynes_2023},
one isomer of the wide family of $\gamma$-graphynes.
These structures can be described as hexagonal benzene rings joined
by the mentioned acetylene molecules or
more precisely acetylenic (triple) bonds,
leading to triangular pores in a completely regular fashion.  Depending on
the number of triple bonds, we can have different compounds with different
triangular nanopores sizes, and hence GDY, contains two acetylene molecules
between the benzenic rings.

Because of the nanoporous regular structure of GDY, one of the many
applications\cite{gr2-rev-fundamental-and-applied,li_graphynes_2023,
  review-gr2-and-applicat-23,zheng_two-dimensional_2023} for which
these structures are specially suited includes
separation at the atomic and molecular
level of different species
\cite{bentley_high-resolution_2022,zhou_gas_2022,griffin_proton_2020,o2-n2-separa-isotop-gr2-qmodel-24}
and, taking advantage of quantum properties,
isotopic separation\cite{sun_limits_2020,c9cp01364d,doi:10.1002/adma.201803772},
being this process in the nanoscale also known as
  {\sl quantum sieving}.  This concept was introduced by Beenakker et al.
  \cite{beenakker-prl-94,BEENAKKER-cpl-95} and the term has become recently
  more popular\cite{quantum-siev-12-rsc}, although the idea of
  quantum contribution to differentiate isotopes was already
  suggested in the earliest 60' of the last century  by
  Freeman\cite{freeman-60-quantum}, commenting of $H_2$ and $D_2$
  adsorption on charcoal at
  low temperature. Isotopes, because they only differ in the
mass, are difficult and costly to separate
by conventional physical or chemical means. They are, however, 
very important in many areas of cutting-edge science, as in the case of
helium\cite{wang_helium_2025} or as a key ingredient in
energy research like hydrogen\cite{rev-30years-istopic-sep-25}.  For
this reason, the advent of new nanoporous 2D materials was seen as a
feasible alternative to traditional methods, for isotopic separation.
Despite the importance of quantum effects, the
approaches to these systems and processes have been carried out
mostly in a classical way.  To recover, at least partially, some of
the quantum behavior at the nanoscale, some attempts have been made, 
mainly by using reduced dimensionality approaches, basically
one dimensional treatments\cite{bartolomei_jpcc:2014,mandra_helium_2014},
using Feynman-Hibbs interaction
potentials\cite{quant-effect-kin-sieving-05,rpbhcgvhb:jpca16}
or other statistical approaches that consider the similitude
between a quantum object and a classical one consisting in
a set of similar particles joined by harmonic
interactions\cite{craig_jcp:2004,craig_jcp_1:2005,rpmd-isotop-he-gr2-2021}.  
But it is precisely for light species
and narrow pores that quantum
effects are more important and thus it is worthy addressing this
problem in a quantum mechanical way, as ``exact'' as possible, given
the difficulties to tackle the large number of particles involved
in these problems.

We presented some ago the first rigorous time-dependent
quantum wave packet approach in three dimensions to
this problem\cite{1st-td3d-quan-iso}, 
where the atomic layer was considered as a rigid membrane
and where the interacting
atom was described fully quantum mechanically.

\begin{figure*}
 \includegraphics[width=\textwidth,height=6.0cm]{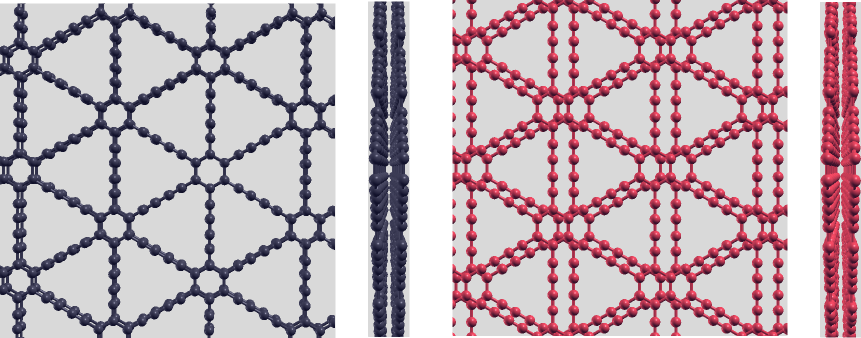} 
      \caption{ Perspective projection of graphdiyne 2D materials. Left: Bilayer
      graphdiyne with an AA  stacking, top and lateral view.  Right: Same bilayer
      material with an AB stacking. (see text) }
  \label{fig-1}
\end{figure*} 

The model was applied to isotopic separation of helium
\cite{1st-td3d-quan-iso,he-isotopic-gr3-wp3d-21} as well as hydrogen
isotopologues\cite{h2-isotopic-gr2-2022-q}.
Results for both helium and hydrogen isotopic separations were very
encouraging, with selectivity values close to $6$, the ratio\cite{permeance:06}
that it is estimated as an acceptable figure for a given material to be
considered in the industry.
The temperatures to get these selectivity values were, however, very low
(below some $\approx 40\; K$) and also the trade-off between selectivities and
fluxes was not very favorable.  Therefore, it would be desirable to explore
new possibilities based on these structures, as  multilayer
2D materials,to explore improvements
  in the sieving performance as well as new possible effects that could
  appear as a consequence of the bilayer structure.
In fact, we are in a situation in which our capacity to virtually
produce many new different materials push the range of applicability
to new scenarios and where theoretical support is very much needed.
In this manner, we can mention the use of doping with several metals
\cite{reider-ijhe-2024,doping-2dmat-17,heteroatomdop-2dma-20,noauthor_doping_2021}, or more interesting to
this work, the ability to produce multilayer
heterostructures\cite{stacking-engi-2dmat-24,stacking-2dmat-21}
in different arrangements.  This development is in the core of new advances,
such as the so-called ``magic angle''
on twisted bilayer graphene\cite{twisted-bilayer-jarillo-2018} or
water desalination technologies.\cite{gr2-multlayer-water-selec-25}
In the latter case GDY was precisely fabricated in an $AA$
stacking tailored to meet the desired separation goal.
In these materials, with some minor discrepancies,
the most stable stacking reported for bilayer
GDY is the AB \cite{nanoscale-12-struct-bi-gr2-gr3}, with an interlayer
distance of 3.40-3.42 \AA, being the AA stacking a bit less stable
(for about 4 meV) and at an interlayer distance of 3.65 \AA $\;$  due to
electron cloud repulsion, both arrangements can be
  seen in Fig.\ref{fig-1}.
Because of the possibilities of
manufacturing these types of bilayer systems we believe it is important
to study the behavior of bilayer GDY structures to understand
new aspects that could arise when an exact quantum approach is used,
for instance, if we can provide a new quantum sieve or there were new
phenomena that could be exploited for these or other applications.

In this paper we have carried out a three dimensional wave packet study of
helium isotopes quantum sieving by
bilayer graphdiyne nanostructures at an energy regime where quantum effects
are very important.  The computed transmission probabilities show a
pattern that is completely different to the monolayer case and that depends on
the geometries of such bilayer material.

\begin{figure}[h!]
      \begin{subfigure}{\textwidth}
        \includegraphics[scale=0.7]{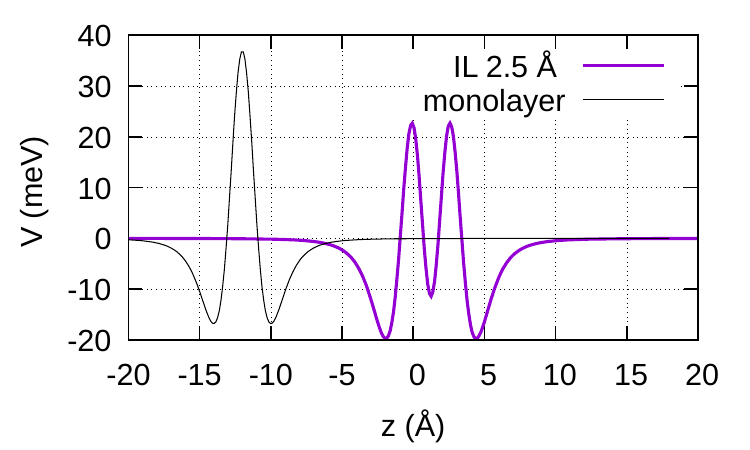}
    \caption{\label{fig2-a}}
   \end{subfigure}
   \begin{subfigure}{\textwidth}
    \includegraphics[scale=0.5]{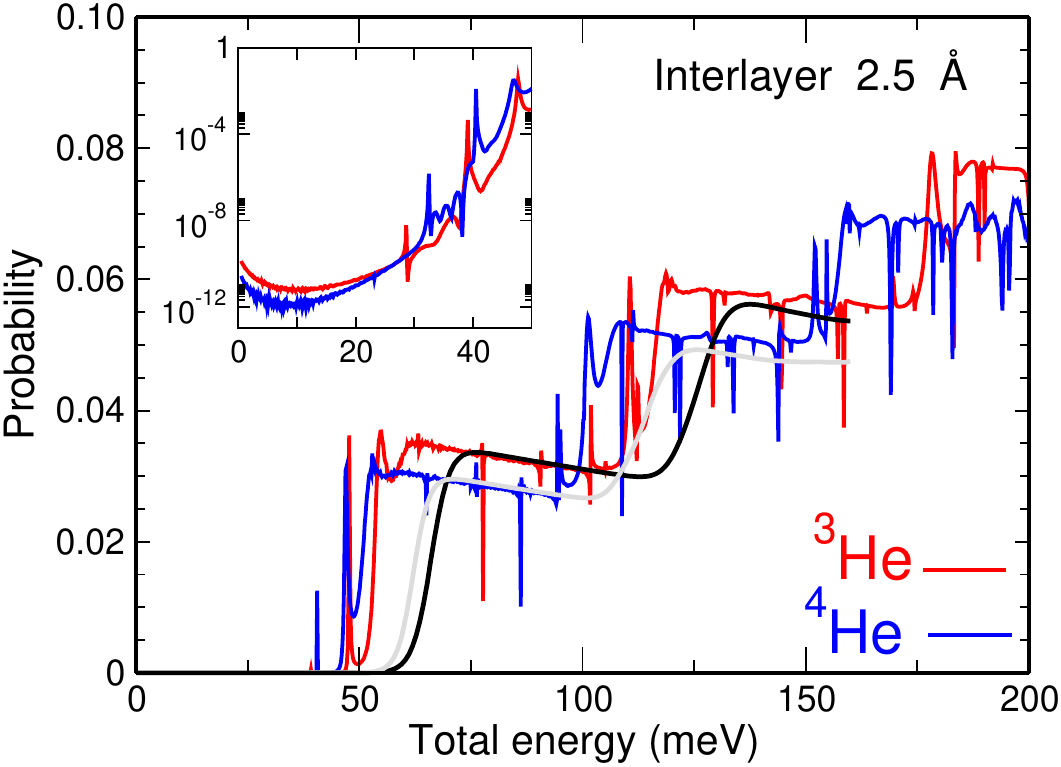}
       \caption{\label{fig2-b}}
   \end{subfigure}
      \caption{ (a)  Interaction potential in a perpendicular approach
       along the $Z$ axis for a bilayer $AA$ nanostructure of interlayer
       separation of  2.5 \text{\AA}. Similar energy profile for the monolayer
      case of Ref.\cite{1st-td3d-quan-iso}, moved to negative values to avoid
      superposition, is included in a thin black line to spot the difference
      with our present case. Note that the sum of two of
        these curves
        in different positions will produce different barriers and wells.
      (b) Probabilities of He transmission, $ P_{trans}(E)$,
        through bilayer graphdiyne,
        $AA$ stacking  when the distance between layers is 2.5 \text{\AA}.
        Insets show the system behavior at very low
        energies. The monolayer graphdiyne case of Ref.\cite{1st-td3d-quan-iso}
        is included as continuous black ($^3He$) and gray ($^4He$) lines as
       a guide to the eye.}
      \label{fig2-in-one}
\end{figure}

\begin{figure}[h]
    \begin{subfigure}{\textwidth}
     \includegraphics[scale=0.7]{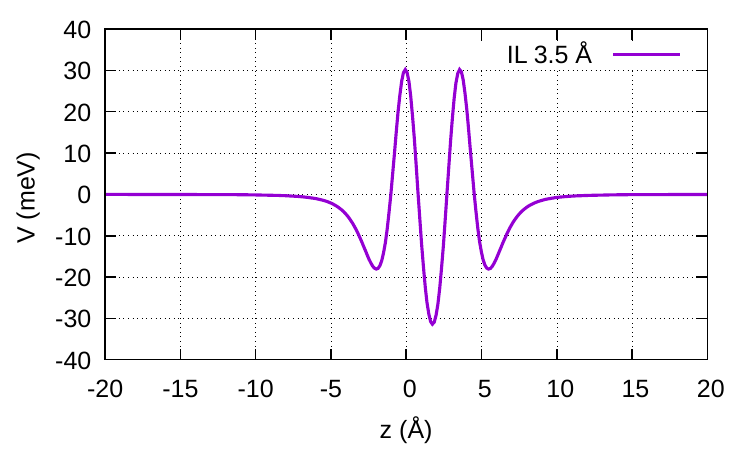}
        \caption{\label{fig3-a}}
    \label{fig3-a}
    \end{subfigure}
       \begin{subfigure}{\textwidth}
     \includegraphics[scale=0.5]{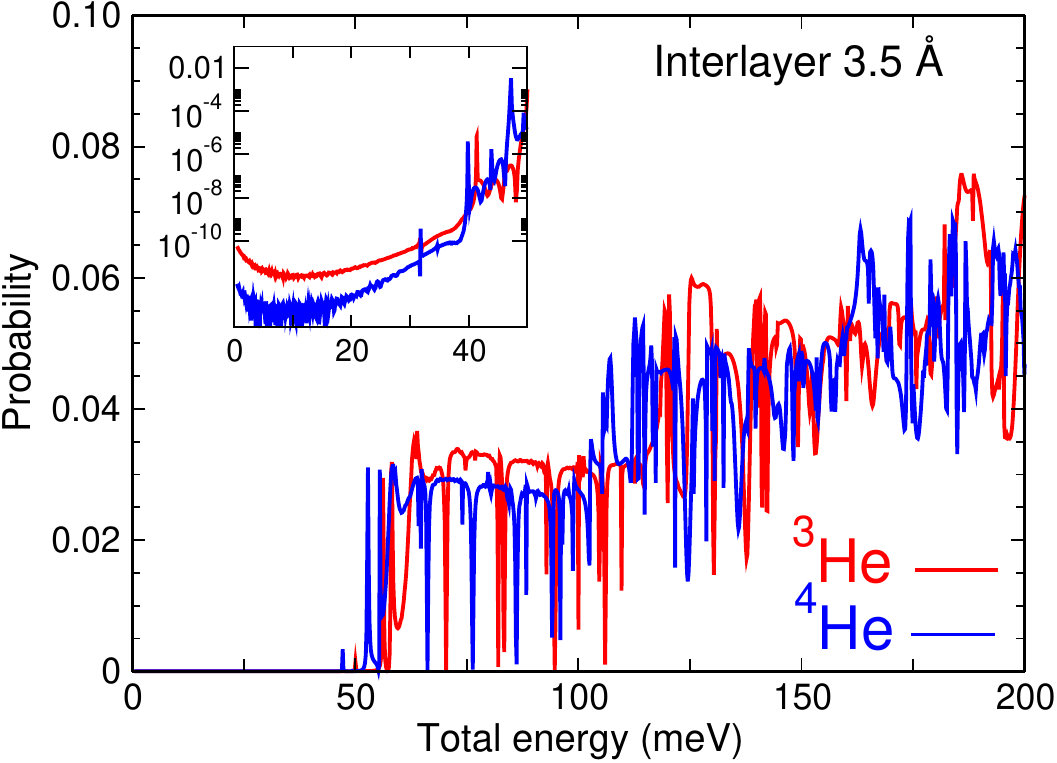}
     \caption{\label{fig3-b}}
    \label{fig3-b}
    \end{subfigure}
       \caption{ (a) Interaction potential in a perpendicular approach
       along the $Z$ axis for a bilayer $AA$ nanostructure of interlayer
       separation of  3.5 \text{\AA}.
       (b) Probabilities of He transmission, $  P_{trans}(E)$,
         through bilayer graphdiyne,
         $AA$ stacking  when the distance between layers is 3.5 5 \text{\AA}.
         Insets show the system behavior at very low
         energies. }
       \label{fig3-in-one}
\end{figure}

\begin{figure}[h]
  \includegraphics[width=8.3cm]{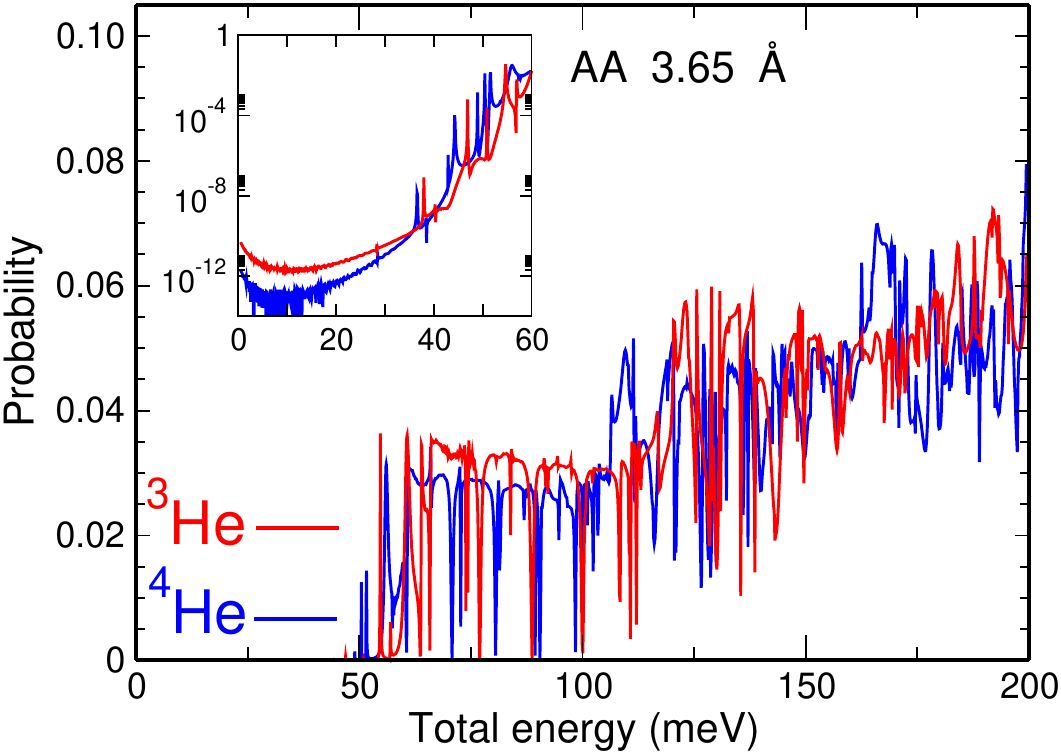}%
  \caption{Probabilities of He transmission through bilayer graphdiyne,
    $AA$ stacking for interlayer separation of 3.65 \text{AA}.
    Inset showing the
    system behavior at low energies. }
  \label{fig4}
\end{figure}

\section{Computational Method}
\label{theory}

In this work, using a similar  model to the one layer GDY
previously mentioned\cite{1st-td3d-quan-iso},  
the Time-Dependent Sch\--r\"odinger Equation,
\begin{equation}
  i \hbar\,\frac{\partial}{\partial t} \Psi(x,y,z,t)=
  [T+V(x,y,z)]\Psi(x,y,z,t)
  \label{tdse}
\end{equation}
\noindent
is employed to propagate a wave packet  on a grid until the
packet completely leaves the interaction
region and is absorbed in a conveniently designed boundary
zone\cite{heather:87,pernot:91}.  The interaction potential 
is constructed as a sum of pairwise interactions, given by an empirical
formula (see Eq. S1) called
Improved Lennard-Jones(ILJ)\cite{ilj} by the authors, and
whose  parameters were
optimized at a high level of theory\cite{bartolomei_jpcc:2014}.
The procedure, briefly sketched, started with
  an annulenic rigid prototype which was optimized within the ``coupled''
  supermolecular
  second-order M{\o}ller-Plesset perturbation theory (MP2C)\cite{mp2c},
  afterwards the ab initio points of the interaction between the prototype and
  a helium atom were used to fine tune the parameters of the mentioned ILJ,
  as it has been explained in our Ref.\cite{bartolomei_jpcc:2014} and
  also in our previous study on graphynes interacting with several
  species\cite{water-gr2-permea-our-14}.   Then, these interactions are
the ones summed up to all the atoms in the
2D membrane to obtain the global 3D interaction potential $V(x,y,z)$,
extended with periodic boundary con\-di\-tions\cite{yinnon:83}.
Once the wave packet is propagated with the Split Operator
method\cite{splitoperator,k&k:1983}, the (total) transmission
probability, which is the sum of the squared transmission
amplitudes, can be obtained from the flux of the stationary wave function
through a surface $z= z_{f}$ separating transmitted from incident and
reflected waves\cite{Miller:74},    
\begin{equation}
  P_{trans}(E)  = \frac{2 \pi \hbar^2}{\mu} \operatorname{Im}
  \left( \int dx dy \,
\Psi^{+*}_E(x,y,z_{f}) \frac{d  \Psi^{+}_E}{dz}\mid_{z=z_{f}} \right).
\label{trprob}
\end{equation}

\noindent
We have employed this flux formula for computing $P_{trans}$, where
$\Psi^{+}_E(x,y,z_{f})$ is obtained from the time-energy Fourier transform
of the evolving wave packet\cite{Zhang:91,cplh2h2:01} and $\hbar$ is
the Planck constant, $\mu$ being the mass of the species crossing
the dividing surface (membrane).


 For a perpendicular incidence, permeances
  (thermal molecular flux per
unit pressure) can be computed as\cite{h2-isotopic-gr2-2022-q} 

\begin{equation}
  S_p(T) = \frac{1}{K_B T} \int_{0}^{\infty} \mathrm{d}v_z\,
  f(v_z, T)\, v_z^2 \mathcal{P}(v_z).
\label{permeation}
\end{equation}

\noindent
where $K_B$ is the Boltzmann constant, $ f(v_z, T)$ the
  velocity  distribution at temperature $T$, and $P(v_z)$ is the
  transmission probability computed with Eq.(\ref{trprob}),
  being $v_z=\sqrt{2E/\mu}$.  The ratio of permeances for two isotopes
is defined as the selectivity.

Parameters and computational details of this study
are given in the Supplementary 
Information (SI).  The bilayer graphdiyne membrane is made up with
two monolayer GDY, based on the unit cell given in Table I,
for the monolayer case and that
  it was optimized as described before.
The transmission probabilities obtained
through Eq.(\ref{trprob}), are computed for several stacking
geometries (Table II) and using the interaction potential previously
deve\-loped\cite{bartolomei_jpcc:2014}.
We assume that interaction between the layers does not
  affect   the dynamical
  quantum calculations, and that electron cloud
  interactions between layers are responsible solely for the stability
  of the stacking
  structures leading to the most stable geometries as it is indicated in
  the next section.
  
\section{Results and discussion}
\label{results}

To characterize the physical phenomena that could appear in a bilayer
heterostructure of graphdiyne,
we have studied 
several stacking arrangements, 
taking advantage of the possibilities offered by the experimental
likelihood\cite{gr2-multlayer-water-selec-25} of virtually producing
almost any geometry. In the  first place we have
  worked out several $AA$ stacking 
  arrangements, with the goal of a better characterization of the features
  appearing in the quantum transport as we change the interlayer
  separation, and finally in a second stage taking the most stable
  stacking geometries ($AA$, $AB$) that lead to the material in Fig.\ref{fig-1},
  to study the corresponding transmission probabilities. In all cases,
  as commented in the previous section, the bilayer GDY is a stacking
  composition of the two monolayer 2D material, that were optimized
previously\cite{bartolomei_jpcc:2014,water-gr2-permea-our-14}. 

The first case considers an $AA$ stacking with a distance between the
layers ($2.5\; \text{\AA}$) closer
than that usually mentioned in the
literature. In this geometry, the presence of
the two layers decreases the barrier to surmount the bilayer system,
with respect to the monolayer case, leading to  a well in between
the two layers as well as slightly deepening
the outside-layers wells as shown in Fig.(\ref{fig2-a}),
that can be compared with the monolayer case\cite{1st-td3d-quan-iso}, that
has been included, on the left, in
the same figure.  Calculated transmission
probabilities are shown in Fig.(\ref{fig2-b}) where we can immediately
appreciate that they keep some of the features
appearing in the monolayer case, such as the influence of tunnel
effect that makes lighter isotopes to cross the filter more easily (as
shown in the inset), while the heavier species dominates
at higher energies in the probability thresholds due to
reduced zero point energy (ZPE) at the transition state\cite{bartolomei_jpcc:2014}.
It can be also noticed several plateaus after a
  sudden probability increasing, that we attributed to the different quasibound
  states within the nanopore in a direction parallel to the layer.
  This behavior can be easily compared with the monolayer
  results since
  for the benefit of the eyes we have included the corresponding probabilities
  in black and gray for the two helium isotopes.  Indeed, it can be noticed
  that the lower barrier produced as
  a consequence of the $AA$ stacking
  is translated into noticeable lower thresholds for transmission.  
   Transmission probabilities are in the base of the
    mass transport through the membrane, as can be seen from
    Eq.\ref{permeation} that provided us with permeances, and thus
    we can already notice that permeances (flux) will be larger.
    In addition, very evident spikes that strikingly contrast with
    the monotonic behavior of the monolayer system,
    appear at regular energy locations,
  which are different depending on the isotope ($^3He/^4He$).
  This feature is a completely new effect that represents a sudden
  increasing or reduction in transmission probabilities,
  superimposed to a staircase-like structure that it is also apparent
  in the monolayer case.  We   attribute this spike-like behavior to quasibound states inside the two layers, as we will discuss in the next paragraphs.

   We show next, in Fig.(\ref{fig3-b}) the transmission
  probabilities for an $AA$ stacking but with interlayer
  separation of $3.5\, \text{\AA}$.  In this case, a similar behavior can be
  observed in the very low energy regime (in the inset of
  Fig.(\ref{fig3-b})), the same plateaus, but now the states (spikes)
  are much more numerous, or in other words, there is a noticeable
  higher density of spikes.
  The effect of the two layers in the energy profile in a direction
  perpendicular to the layer, as it is shown in Fig.(\ref{fig3-a}), is
  in this case to increase the barrier height with respect to the
  previous case (but still lower than
  the monolayer membrane,
  cf. Fig.(\ref{fig2-a})) leading at the
  same time to a much deeper
  well in between layers. This will mean, because of
    the lower threshold, larger
    permeances than in the monolayer case, although smaller than
    in the previous case, and larger number of 
    spikes due to the deeper well between the layers. We have to attribute
  again the large congestion of these spikes,  these abrupt
  changes in transmission probabilities, to the quasibound states
  that can be formed inside the layers of the GDY
  heterostructure, that in a very crude model of a  parallelepiped box
  will contain energy levels with separations inversely
  proportional to their dimensions.  The dimension that changes in the
  two previous cases is the interlayer separation,
  therefore in our second $AA$ stacking case we will see a larger
  number of energy levels because of this increased separation.

   Turning to a more realistic case, $AA$ stacking with an interlayer
  separation of 3.65 $\text{\AA}$, i.e. a slight increase of $0.15\; \text{\AA}$
  to the previously studied layer position, we can observe
  in Fig.(\ref{fig4}) that the probabilities follow the previous pattern
  with an increased density of spikes that makes difficult to differentiate
  between both helium isotopes, although still there are clear regions where
  the probability of one of the isotopes is predominant over the other. This
  feature can be better observed taking a closer look to the probabilities in a
  narrower energy range (see S.I. Fig. 10), where it can be seen the clear
  changes in thresholds and maxima and minima with evident differences
  between both isotopes.

    From the previous discussion, we stress that the change in
    the interlayer distance leads
    to a different number and positions of maxima and minima,
  with a gradual increment in these characteristics when we move from
  the shorter ($2.5\;  \text{\AA}) $ to the larger ($3.65\;  \text{\AA}) $  
  interlayer separation (as it can be appreciated in S.I. Fig.11).
  
\begin{figure}[h]
  \includegraphics[width=8.3cm]{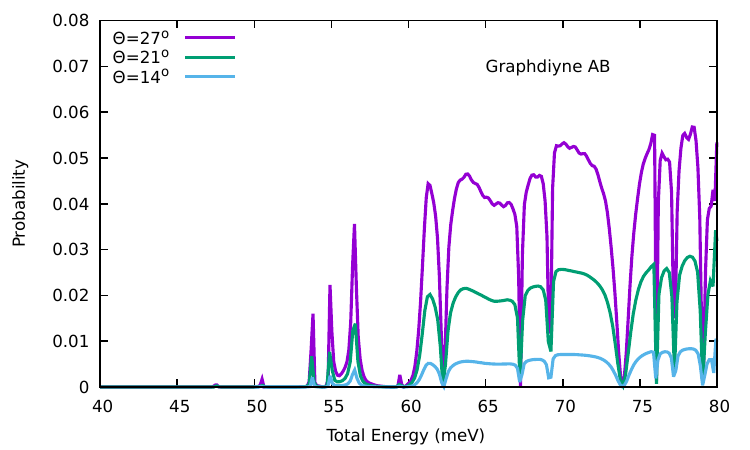}
  \caption{Probabilities of $^3$He transmission through bilayer graphdiyne
    for $AB$ stacking.  Label $\theta$ indicates the central angle of
    incidence in the plane $XZ$, for an increasing initial total energy.  }
  \label{fig5}
\end{figure}

 To further confirm the origin of the probabilities behavior, and
  the influence of the interaction on these magnitudes, we have carried out
  another calculation with an $AB$ stacking, borrowing the
  parameters found in Ref.\cite{applmaterours:2016,apriliyanto_multilayer_2022}
  (S.I., Table II).   In this case a
  perpendicular approach of helium atoms will result into
  a vanishing transmission
  since the nanopores of the two layers are no longer aligned
  with respect to each other.  
  There is however, a minimum energy path connecting the
  centers of the pores of the two layers, that follows a
  straight line.  In this direction, corresponding approximately to
  $\theta \approx 27^{0}$ in the $XZ$ plane, the energy profile
  (Fig. S11, in the S.I.) is similar to the previously
  commented for the $AA$ stacking, and correspondingly
  the transmission probabilities are larger and similar to the
  previous $AA$ stacking cases.  Moreover, calculations at other
  incident angles, different from that following the minimum energy
  path, indicate that the probabilities are smaller, as it can
  be appreciated in Fig.(\ref{fig5}), with
  the spikes features in the same positions for all incident angles, confirming
  once again that the behavior depends on the heterostructure geometry
  and not on other peculiarities of the system, for example the
  selective adsorption resonances formed  on
  the surface\cite{lifetime-he-surf-td-94}
  that can be observed at very low
  incident angle or barrierless nanoporous systems
  \cite{1st-td3d-quan-iso,he-isotopic-gr3-wp3d-21}.

\begin{figure}[h]
  \includegraphics[width=8.3cm]{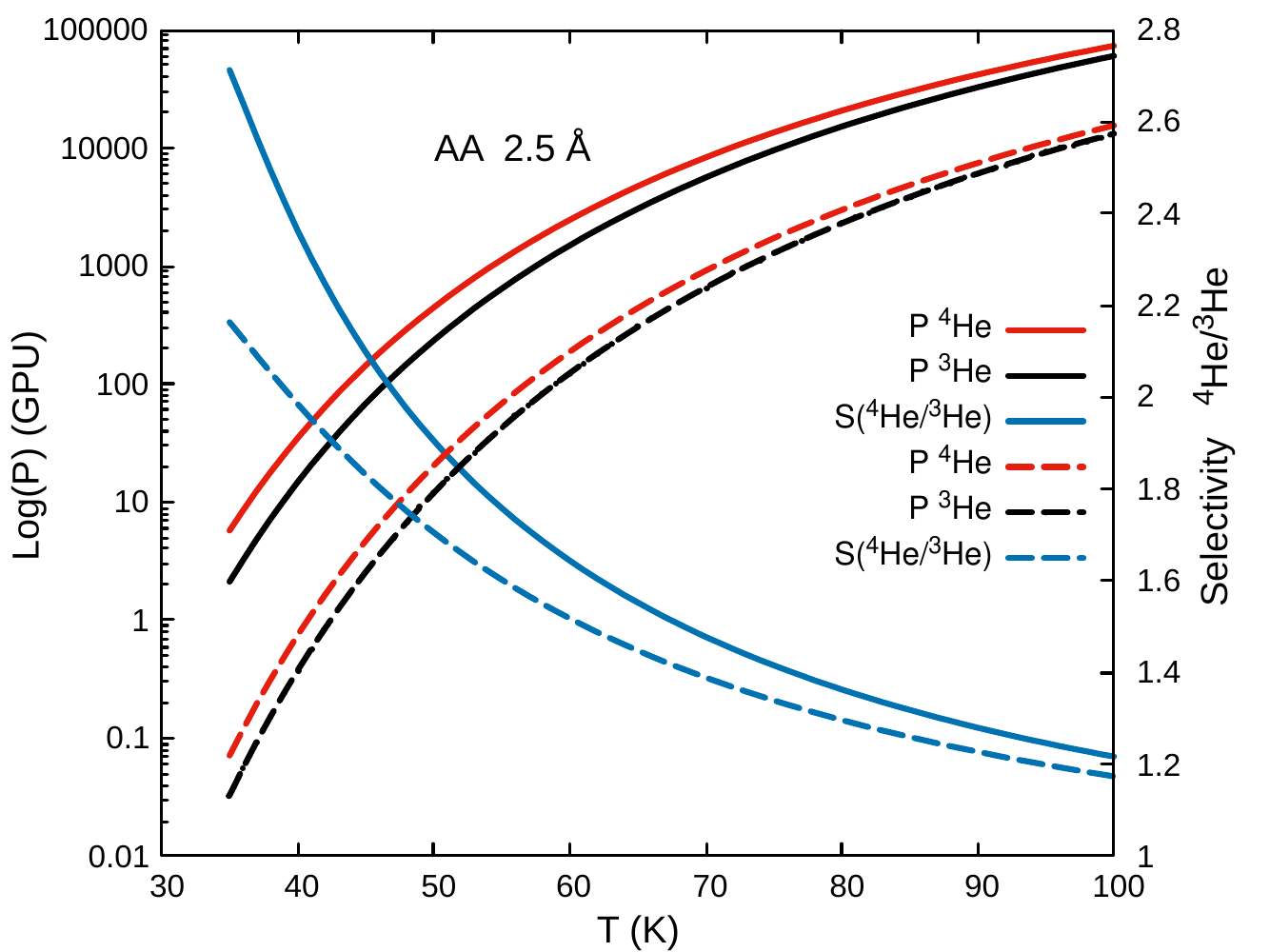}%
  \caption{Permeances and selectivities.  Continuous line for bilayer
    graphdiyne with $AA$ stacking at $2.5\; \text{\AA} $, dashed lines for the
    monolayer case of Ref.\cite{1st-td3d-quan-iso}. Selectivities
    $S(^4He/^3He)$ in blue color, continuous line for bilayer and
    dashed line for the monolayer case. Permeances are
    in GPU (Gas Permeation Unit), where
    $1\; GPU = 3.35\; \times \; 10^{-10} mol\;  m^{-2} s^{-1} Pa^{-1}$, and
  $m,\, s,\,Pa$ stand for meter, second and pascal. }
  \label{fig6}
\end{figure}

\begin{figure}[h]
  \includegraphics[width=8.3cm]{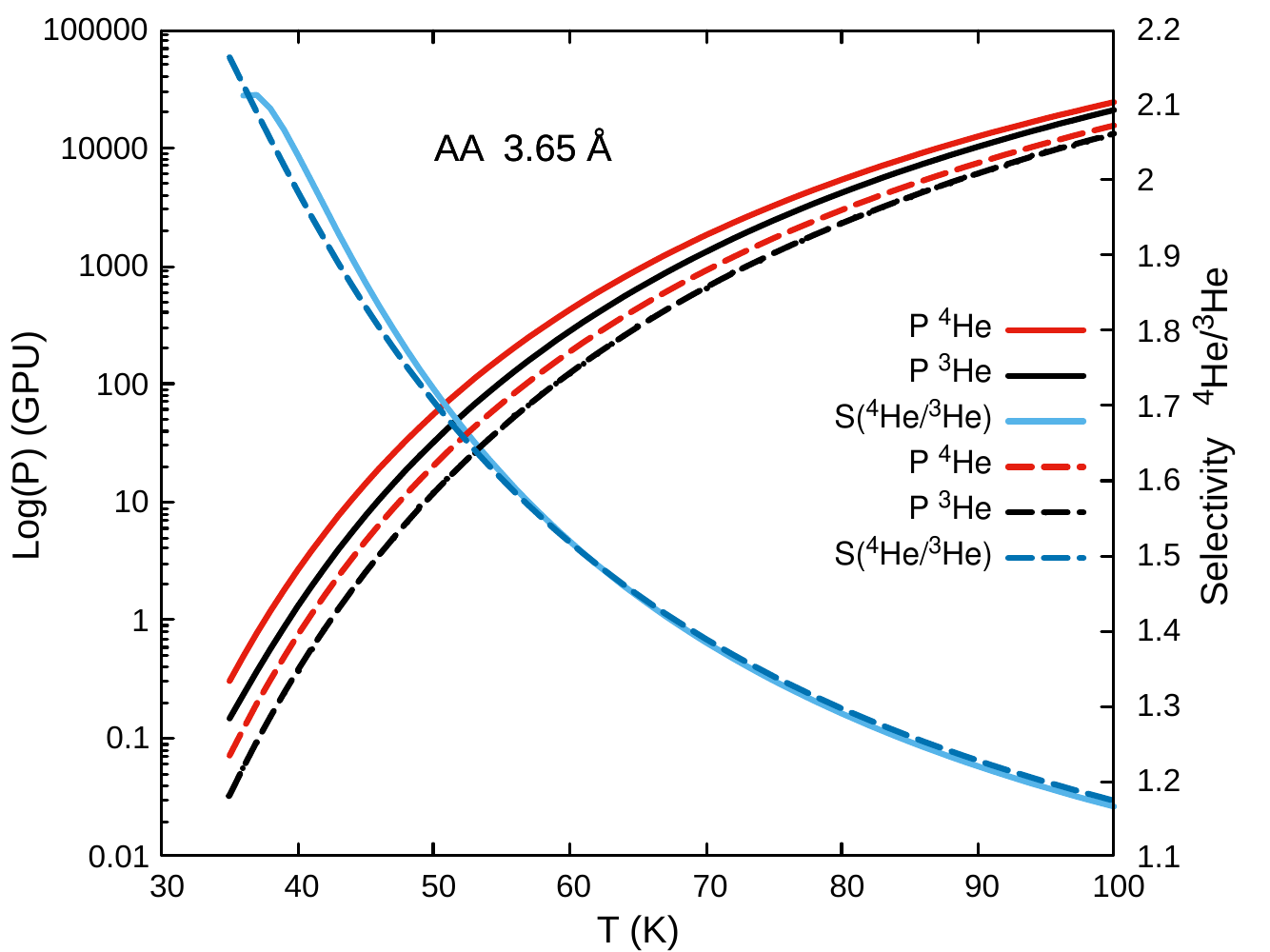}%
  \caption{Permeances and selectivities. Same as in Fig.\ref{fig4}
    for the case of $AA$ stacking at $3.65\; \text{\AA} $. }
  \label{fig7}
\end{figure}

    Therefore, we find that a bilayer 2D structure
    will enhance transmission, hence 
  increasing permeances, and it will show a particular pattern of maxima and
  minima in the probabilities, given by Eq.\ref{trprob}, that can also
  affect transmission.  The density of spikes increases
  with the interlayer separation leading to a practical overlap of maxima and
  minima in very narrow energy windows that will imply more difficulties
  to distinguish between different isotopes. In order to show the
  mentioned enhancement in transmission, we have computed permeances as
  it was shown using Eq.\ref{permeation}.  In Fig.(\ref{fig6}) we show permeances
  for the smallest stacking distance and as we have already discussed
  there is a large difference, of $\approx 2$ orders of magnitude, between monolayer
  and bilayer membranes permeances.  In the same panel we can also appreciate that
  there is an improvement in $^4He/^3He$ selectivity albeit it is not large.
  For the
  case of $AA$ stacking at $3.65\; \text{\AA}$, we can see in Fig.\ref{fig7}
  that there is a small improvement in permeances, however
  selectivities are very similar to the monolayer case, because the high
  congestion of spikes makes more difficult to
  distinguish between isotopes giving a performance similar to
  the monolayer case.

 \section{Conclusions}
 As a word of conclusion we have shown a new resonant effect appearing
 in multilayer 2D materials, specifically in bilayer graphdiyne, and
 shown that large spikes at different energies appear superimposed to
 the regular pattern of transmission probabilities that we could find
 in a monolayer system.  These properties could be used in isotopic
 separation by tailoring the appropriate heterostructure and a
 gaseous beam of adequate energy, although for the
   more stable stacking the high density of spikes makes  difficult
   to isolate regions of stronger predominance of one
   isotope over the other.  Much more important is the enhancement
   effect that can be obtained in the quantum transport, or
   permeances.  The features depend on the
 geometry of the structure, modifying a pattern mostly given
 by the monolayer 2D material.  This effect has not been shown before in a
 three dimensional quantum study, and only a similar case
 has been reported in one
 dimensional systems\cite{mandra_helium_2014} where, because of the
 one dimensional character of the physical problem, the resonances
 formed lead to probabilities close to unity, while
 in the present three-dimensional case the probabilities exhibit stair-like
 shapes with superimposed strong spikes at different energy positions. It is the
 potential barriers at the nanopores that produces a given shape of
 the transmission probabilities on top of which the resonant behavior is superimposed
 as strong spikes at different energy positions.
 The features here
 shown present also some resemblances with known resonant tunneling
 phenomena that appears in electronics in semiconductor structures, but with
 the peculiarity that here they are atoms, not electrons, the ones that
 can resonantly increase or decrease their transmission due to
 quasibound states in the space between the nanopores  layers.
 This property could be used to enhance quantum sieving and separations.
 The intercalation of different species into
 2D materials\cite{yang_intercalation_2024} could also be engineered
 to serve in several research fields such as
 batteries\cite{recent-advan-2dmat-batteries-24,gr2-baterias-max-24}
 and other cases were the insertion and intercalation of atoms and ions
 could be of interest.



\section{Acknowledgments}
This work has been supported by  Agencia Estatal de Investigaci\'on,
  Ministerio de Ciencia, Innovaci\'on y Universidades, SPAIN.
  (Project ref. PID2023-149406NB-I00 supported by
  MCI/AEI/10.13039/5501100011003/FEDER, EU).  CESGA is also acknowledged
  for allocation of computing time.





\bibliographystyle{apsrev4-2}

\bibliography{gr2-bilayer}

@STRING{jpca	= {J. Phys. Chem. A} }

@Article{	  1st-td3d-quan-iso,
  issn		= {1932-7447},
  journal	= {J. Phys. Chem. C},
  pages		= {19751--19757},
  volume	= {121},
  year		= {2017},
  title		= {Wave Packet Calculations of the Quantum Transport of Atoms
		  Through Nanoporous Membranes},
  author	= {A. Gij{\'o}n and J. Campos-Mart{\'\i}nez and M.I.
		  Hern{\'a}ndez}
}

@Article{	  applmaterours:2016,
  author	= { M. Bartolomei and G. Giorgi },
  title		= { a Novel Nanoporous Graphite Based on Graphynes:
		  First-Principles Structure and Carbon Dioxide Preferential
		  Physisorption },
  journal	= { ACS Appl. Mater. Interfaces },
  pages		= { 27996-28003 },
  year		= 2016,
  volume	= 8
}

@Article{	  apriliyanto_multilayer_2022,
  title		= {Multilayer {Graphtriyne} {Membranes} for {Separation} and
		  {Storage} of {CO2}: {Molecular} {Dynamics} {Simulations} of
		  {Post}-{Combustion} {Model} {Mixtures}},
  volume	= {27},
  issn		= {1420-3049},
  shorttitle	= {Multilayer {Graphtriyne} {Membranes} for {Separation} and
		  {Storage} of {CO2}},
  url		= {https://www.mdpi.com/1420-3049/27/18/5958},
  doi		= {10.3390/molecules27185958},
  language	= {en},
  number	= {18},
  urldate	= {2022-12-20},
  journal	= {Molecules},
  author	= {Apriliyanto, Yusuf Bramastya and Faginas-Lago, Noelia and
		  Evangelisti, Stefano and Bartolomei, Massimiliano and
		  Leininger, Thierry and Pirani, Fernando and Pacifici,
		  Leonardo and Lombardi, Andrea},
  month		= jan,
  year		= {2022},
  pages		= {5958}
}

@Article{	  bartolomei_jpcc:2014,
  author	= {Bartolomei, Massimiliano and Carmona-Novillo, Estela and
		  Hern{\'{a}}ndez, Marta I. and Campos-Mart{\'{i}}nez,
		  Jos{\'{e}} and Pirani, Fernando and Giorgi, Giacomo},
  doi		= {10.1021/jp510124e},
  issn		= {19327455},
  journal	= {J. Phys. Chem. C},
  number	= {51},
  pages		= {29966--29972},
  title		= {{Graphdiyne pores: Ad hoc openings for helium separation
		  applications}},
  volume	= {118},
  year		= {2014}
}

@Article{	  baughman_structureproperty_gr-1987,
  author	= {Baughman, R. H. and Eckhardt, H. and Kertesz, M.},
  title		= {Structure‐property predictions for new planar forms of
		  carbon: {Layered} phases containing sp2 and sp atoms},
  journal	= {The Journal of Chemical Physics},
  year		= {1987},
  volume	= {87},
  number	= {11},
  pages		= {6687--6699},
  month		= dec,
  issn		= {0021-9606},
  doi		= {10.1063/1.453405},
  shorttitle	= {Structure‐property predictions for new planar forms of
		  carbon},
  urldate	= {2024-01-12}
}

@Article{	  bentley_high-resolution_2022,
  author	= {Bentley, Cameron L. and Kang, Minkyung and Bukola, Saheed
		  and Creager, Stephen E. and Unwin, Patrick R.},
  title		= {High-{Resolution} {Ion}-{Flux} {Imaging} of {Proton}
		  {Transport} through {Graphene}{\textbar}{Nafion}
		  {Membranes}},
  journal	= {ACS Nano},
  year		= {2022},
  volume	= {16},
  number	= {4},
  pages		= {5233--5245},
  month		= apr,
  issn		= {1936-0851},
  doi		= {10.1021/acsnano.1c05872},
  urldate	= {2022-06-02}
}

@Article{	  c9cp01364d,
  author	= {Motallebiour, Maryam S. and Karimi-Sabet, Javad and
		  Maghari, Ali},
  title		= {4He/3He separation using oxygen-functionalized nanoporous
		  graphene},
  journal	= {Phys. Chem. Chem. Phys.},
  year		= {2019},
  volume	= {21},
  pages		= {12414-12422},
  doi		= {10.1039/C9CP01364D},
  issue		= {23}
}

@Article{	  cplh2h2:01,
  author	= {D. di Domenico and M. I. Hern{\'a}ndez and J.
		  Campos-Mart{\'\i}nez},
  title		= {A Time-Dependent Wave Packet Approach for Reaction and
		  Dissociation in H$_2$+H$_2$},
  journal	= {Chem. Phys. Lett.},
  pages		= {177-184},
  year		= 2001,
  volume	= 342
}

@Article{	  craig_jcp:2004,
  author	= {Craig, Ian R. and Manolopoulos, David E.},
  doi		= {10.1063/1.1777575},
  issn		= {00219606},
  journal	= {J. Chem. Phys.},
  number	= {8},
  pages		= {3368--3373},
  title		= {{Quantum statistics and classical mechanics: Real time
		  correlation functions from ring polymer molecular
		  dynamics}},
  volume	= {121},
  year		= {2004}
}

@Article{	  craig_jcp_1:2005,
  author	= {Craig, Ian R. and Manolopoulos, David E.},
  doi		= {10.1063/1.1954769},
  issn		= {00219606},
  journal	= {J. Chem. Phys.},
  number	= {3},
  pages		= {034102},
  title		= {{A refined ring polymer molecular dynamics theory of
		  chemical reaction rates}},
  volume	= {123},
  year		= {2005}
}

@Article{	  doi:10.1002/adma.201803772,
  author	= {Qiu, Hu and Xue, Minmin and Shen, Chun and Zhang, Zhuhua
		  and Guo, Wanlin},
  title		= {Graphynes for Water Desalination and Gas Separation},
  journal	= {Advanced Materials},
  year		= {2019},
  volume	= {31},
  number	= {42},
  pages		= {1803772},
  doi		= {10.1002/adma.201803772},
  eprint	= {https://onlinelibrary.wiley.com/doi/pdf/10.1002/adma.201803772}
}

@Article{	  gr-rev-syn-prop-app-23-flatchem,
  author	= {Garima Narang and Divyam Bansal and Shaina Joarder and
		  Prashant Singh and Loveneesh Kumar and Vivek Mishra and
		  Sangeeta Singh and Kaniki Tumba and Kamlesh Kumari},
  title		= {A review on the synthesis, properties, and applications of
		  graphynes},
  journal	= {FlatChem},
  year		= {2023},
  volume	= {40},
  pages		= {100517},
  issn		= {2452-2627},
  doi		= {10.1016/j.flatc.2023.100517},
  url		= {https://www.sciencedirect.com/science/article/pii/S2452262723000491}
}

@Article{	  gr2-review-22,
  author	= {Fang, Yan and Liu, Yuxin and Qi, Lu and Xue, Yurui and Li,
		  Yuliang},
  title		= {2D graphdiyne: an emerging carbon material},
  journal	= {Chem. Soc. Rev.},
  year		= {2022},
  volume	= {51},
  pages		= {2681-2709},
  doi		= {10.1039/D1CS00592H},
  issue		= {7}
}

@Article{	  griffin_proton_2020,
  author	= {Griffin, Eoin and Mogg, Lucas and Hao, Guang-Ping and
		  Kalon, Gopinadhan and Bacaksiz, Cihan and Lopez-Polin,
		  Guillermo and Zhou, T.y. and Guarochico, Victor and Cai,
		  Junhao and Neumann, Christof and Winter, Andreas and Mohn,
		  Michael and Lee, Jong Hak and Lin, Junhao and Kaiser, Ute
		  and Grigorieva, Irina V. and Suenaga, Kazu and Özyilmaz,
		  Barbaros and Cheng, Hui-Min and Ren, Wencai and Turchanin,
		  Andrey and Peeters, Francois M. and Geim, Andre K. and
		  Lozada-Hidalgo, Marcelo},
  title		= {Proton and {Li}-{Ion} {Permeation} through {Graphene} with
		  {Eight}-{Atom}-{Ring} {Defects}},
  journal	= {ACS Nano},
  year		= {2020},
  volume	= {14},
  number	= {6},
  pages		= {7280--7286},
  month		= jun,
  issn		= {1936-0851},
  note		= {Publisher: American Chemical Society},
  doi		= {10.1021/acsnano.0c02496},
  urldate	= {2024-01-25}
}

@Article{	  h2-isotopic-gr2-2022-q,
  author	= {García-Arroyo, Esther and Campos-Martínez, José and
		  Bartolomei, Massimiliano and Pirani, Fernando and
		  Hernández, Marta I.},
  title		= {Molecular hydrogen isotope separation by a graphdiyne
		  membrane: a quantum-mechanical study},
  journal	= {Phys. Chem. Chem. Phys.},
  year		= {2022},
  volume	= {24},
  pages		= {15840-15850},
  doi		= {10.1039/D2CP01044E},
  issue		= {26}
}

@Article{	  heather:87,
  author	= {Heather, Robert and Metiu, Horia},
  title		= {An efficient procedure for calculating the evolution of
		  the wave function by fast Fourier transform methods for
		  systems with spatially extended wave function and localized
		  potential},
  journal	= {The Journal of Chemical Physics},
  volume	= {86},
  number	= {9},
  pages		= {5009-5017},
  year		= {1987},
  month		= {05},
  issn		= {0021-9606},
  doi		= {10.1063/1.452672},
  url		= {https://doi.org/10.1063/1.452672}
}

@Article{	  k&k:1983,
  author	= { D. Kosloff and R. Kosloff},
  title		= {A Fourier Method Solution for the Time Dependent
		  Schr{\"o}dinger Equation as a Tool in Molecular Dynamics },
  journal	= {J. Comp. Phys.},
  pages		= {35-53},
  year		= 1983,
  volume	= 52
}

@Article{	  kang_graphyne_2019,
  author	= {Kang, Jun and Wei, Zhongming and Li, Jingbo},
  title		= {Graphyne and {Its} {Family}: {Recent} {Theoretical}
		  {Advances}},
  journal	= {ACS Applied Materials \& Interfaces},
  year		= {2019},
  volume	= {11},
  number	= {3},
  pages		= {2692--2706},
  month		= jan,
  issn		= {1944-8244},
  doi		= {10.1021/acsami.8b03338},
  shorttitle	= {Graphyne and {Its} {Family}},
  urldate	= {2020-11-11}
}

@Article{	  li_architecture_2010,
  title		= {Architecture of graphdiyne nanoscale films},
  volume	= {46},
  issn		= {1364-548X},
  url		= {https://pubs.rsc.org/en/content/articlelanding/2010/cc/b922733d},
  doi		= {10.1039/B922733D},
  language	= {en},
  number	= {19},
  urldate	= {2024-01-22},
  journal	= {Chemical Communications},
  author	= {Li, Guoxing and Li, Yuliang and Liu, Huibiao and Guo,
		  Yanbing and Li, Yongjun and Zhu, Daoben},
  month		= may,
  year		= {2010},
  pages		= {3256--3258}
}

@Article{	  li_graphynes_2023,
  author	= {Li, Hao and Lim, Jong Hyeon and Lv, Yipin and Li, Nannan
		  and Kang, Baotao and Lee, Jin Yong},
  title		= {Graphynes and {Graphdiynes} for {Energy} {Storage} and
		  {Catalytic} {Utilization}: {Theoretical} {Insights} into
		  {Recent} {Advances}},
  journal	= {Chemical Reviews},
  year		= {2023},
  volume	= {123},
  number	= {8},
  pages		= {4795--4854},
  month		= apr,
  issn		= {0009-2665},
  note		= {Publisher: American Chemical Society},
  doi		= {10.1021/acs.chemrev.2c00729},
  shorttitle	= {Graphynes and {Graphdiynes} for {Energy} {Storage} and
		  {Catalytic} {Utilization}},
  urldate	= {2024-09-11}
}

@Article{	  mandra_helium_2014,
  author	= {Mandrà, Salvatore and Schrier, Joshua and Ceotto,
		  Michele},
  title		= {Helium {Isotope} {Enrichment} by {Resonant} {Tunneling}
		  through {Nanoporous} {Graphene} {Bilayers}},
  journal	= {The Journal of Physical Chemistry A},
  year		= {2014},
  volume	= {118},
  number	= {33},
  pages		= {6457--6465},
  month		= aug,
  issn		= {1089-5639},
  doi		= {10.1021/jp502548r},
  urldate	= {2022-09-15}
}

@Article{	  o2-n2-separa-isotop-gr2-qmodel-24,
  author	= {Rafiei, Maryam A. and Campos-Martínez, José and
		  Bartolomei, Massimiliano and Pirani, Fernando and Maghari,
		  Ali and Hernández, Marta I.},
  title		= {Separation of oxygen from nitrogen using a graphdiyne
		  membrane: a quantum-mechanical study},
  journal	= {Phys. Chem. Chem. Phys.},
  year		= {2024},
  volume	= {26},
  number	= {37},
  pages		= {24553--24563},
  note		= {Publisher: The Royal Society of Chemistry},
  doi		= {10.1039/D4CP02287D}
}

@Article{	  pan_graphdiyne_2021,
  author	= {Pan, Chuanqi and Wang, Chenyang and Fang, Yarong and Zhu,
		  Yuhua and Deng, Hongtao and Guo, Yanbing},
  title		= {Graphdiyne: an emerging two-dimensional ({2D}) carbon
		  material for environmental remediation},
  journal	= {Environmental Science: Nano},
  year		= {2021},
  month		= may,
  issn		= {2051-8161},
  doi		= {10.1039/D1EN00231G},
  language	= {en},
  shorttitle	= {Graphdiyne},
  url		= {https://pubs.rsc.org/en/content/articlelanding/2021/en/d1en00231g},
  urldate	= {2021-07-08}
}

@Article{	  permeance:06,
  author	= { Z. Zhou },
  title		= { Permeance Should Be Used to Characterize the Productivity
		  of a Polymeric Gas Separation Membrane },
  journal	= {J. Membr. Sci.},
  pages		= {754-756},
  year		= 2006,
  volume	= 281
}

@Article{	  pernot:91,
  author	= {P. Pernot and W. A. Lester},
  title		= {Multidimensional Wave-Packet Analysis: Splitting Method
		  for Time-Resolved Property Determination},
  journal	= {Int. J. Quantum Chem.},
  pages		= {577-588},
  year		= 1991,
  volume	= 40
}

@Article{	  rpbhcgvhb:jpca16,
  title		= {Examination of the {Feynman-Hibbs} Approach in the Study
		  of {Ne$_N$}-Coronene Clusters at Low Temperatures},
  author	= {Rodr\'{\i}guez-Cantano, Roc\'{\i}o and P\'erez de Tudela,
		  Ricardo and Bartolomei, Massimiliano and Hern\'andez, Marta
		  I. and Campos-Mart\'{\i}nez, Jos\'e and Gonz\'alez-Lezana,
		  Tom\'as and Villarreal, Pablo and Hern\'andez-Rojas, Javier
		  and Bret\'on, Jos\'e},
  journal	= jpca,
  year		= {2016},
  number	= {27},
  pages		= {5370-5379},
  volume	= {120},
  doi		= {10.1021/acs.jpca.6b01926}
}

@Article{	  rpmd-isotop-he-gr2-2021,
  author	= {Bhowmick, Somnath and Hernández, Marta I. and
		  Campos-Martínez, José and Suleimanov, Yury V.},
  title		= {Isotopic separation of helium through graphyne membranes:
		  a ring polymer molecular dynamics study},
  journal	= {Phys. Chem. Chem. Phys.},
  year		= {2021},
  volume	= {23},
  number	= {34},
  pages		= {18547--18557},
  note		= {Publisher: The Royal Society of Chemistry},
  doi		= {10.1039/D1CP02121D}
}

@Article{	  splitoperator,
  author	= { M. D. Feit and J. A. Fleck and A. Steiger},
  title		= {Solution of the Schr{\"o}dinger Equation by a Spectral
		  Method},
  journal	= {J. Comput. Phys.},
  pages		= {412-433},
  year		= 1982,
  volume	= 47
}

@Article{	  sun_limits_2020,
  title		= {Limits on gas impermeability of graphene},
  volume	= {579},
  issn		= {1476-4687},
  url		= {https://www.nature.com/articles/s41586-020-2070-x},
  doi		= {10.1038/s41586-020-2070-x},
  language	= {en},
  number	= {7798},
  urldate	= {2024-01-25},
  journal	= {Nature},
  author	= {Sun, P. Z. and Yang, Q. and Kuang, W. J. and Stebunov, Y.
		  V. and Xiong, W. Q. and Yu, J. and Nair, R. R. and
		  Katsnelson, M. I. and Yuan, S. J. and Grigorieva, I. V. and
		  Lozada-Hidalgo, M. and Wang, F. C. and Geim, A. K.},
  month		= mar,
  year		= {2020},
  note		= {Number: 7798 Publisher: Nature Publishing Group},
  pages		= {229--232}
}

@Article{	  xu_transparent_2019,
  author	= {Xu, Jiyu and Jiang, Hongyu and Shen, Yutian and Li,
		  Xin-Zheng and Wang, E. G. and Meng, Sheng},
  title		= {Transparent proton transport through a two-dimensional
		  nanomesh material},
  journal	= {Nature Communications},
  year		= {2019},
  volume	= {10},
  number	= {1},
  pages		= {3971},
  month		= sep,
  issn		= {2041-1723},
  doi		= {10.1038/s41467-019-11899-y},
  language	= {en},
  url		= {https://www.nature.com/articles/s41467-019-11899-y},
  urldate	= {2021-09-28}
}

@Article{	  yasuda_efficient_2022,
  author	= {Yasuda, Satoshi and Matsushima, Hisayoshi and Harada,
		  Kenji and Tanii, Risako and Terasawa, Tomo-o and Yano,
		  Masahiro and Asaoka, Hidehito and Gueriba, Jessiel Siaron
		  and Diño, Wilson Agerico and Fukutani, Katsuyuki},
  title		= {Efficient {Hydrogen} {Isotope} {Separation} by {Tunneling}
		  {Effect} {Using} {Graphene}-{Based} {Heterogeneous}
		  {Electrocatalysts} in {Electrochemical} {Hydrogen}
		  {Isotope} {Pumping}},
  journal	= {ACS Nano},
  year		= {2022},
  volume	= {16},
  number	= {9},
  pages		= {14362--14369},
  month		= sep,
  issn		= {1936-0851},
  doi		= {10.1021/acsnano.2c04655},
  urldate	= {2022-10-05}
}

@Article{	  yinnon:83,
  author	= { A. T. Yinnon and R. Kosloff},
  title		= {A Quantum-Mechanical Time-Dependent Simulation of the
		  Scattering from a Stepped Surface },
  journal	= {Chem. Phys. Lett.},
  pages		= {216-223},
  year		= 1983,
  volume	= 102
}

@Article{	  zhang:91,
  author	= {Zhang, Dong H. and Zhang, John Z. H.},
  title		= {Full‐dimensional time‐dependent treatment for
		  diatom–diatom reactions: The H2+OH reaction},
  journal	= {The Journal of Chemical Physics},
  volume	= {101},
  number	= {2},
  pages		= {1146-1156},
  year		= {1994},
  month		= {07},
  issn		= {0021-9606},
  doi		= {10.1063/1.467808},
  url		= {https://doi.org/10.1063/1.467808}
}

@Article{	  zhou_gas_2022,
  title		= {Gas permeation through graphdiyne-based nanoporous
		  membranes},
  volume	= {13},
  issn		= {2041-1723},
  url		= {https://www.nature.com/articles/s41467-022-31779-2},
  doi		= {10.1038/s41467-022-31779-2},
  language	= {en},
  number	= {1},
  urldate	= {2023-03-15},
  journal	= {Nature Communications},
  author	= {Zhou, Zhihua and Tan, Yongtao and Yang, Qian and Bera,
		  Achintya and Xiong, Zecheng and Yagmurcukardes, Mehmet and
		  Kim, Minsoo and Zou, Yichao and Wang, Guanghua and
		  Mishchenko, Artem and Timokhin, Ivan and Wang, Canbin and
		  Wang, Hao and Yang, Chongyang and Lu, Yizhen and Boya,
		  Radha and Liao, Honggang and Haigh, Sarah and Liu, Huibiao
		  and Peeters, Francois M. and Li, Yuliang and Geim, Andre K.
		  and Hu, Sheng},
  year		= {2022},
  pages		= {4031}
}

@Article{	  review-gr2-and-applicat-23,
  author	= {Li, Xiaotong and Cui, Xinwei and Zhang, Ling and Du,
		  Jiang},
  title		= {Review of Graphdiyne-Based Nanostructures and Their
		  Applications},
  journal	= {ACS Applied Nano Materials},
  year		= {2023},
  volume	= {6},
  number	= {22},
  pages		= {20493-20522},
  doi		= {10.1021/acsanm.3c03038}
}

@Article{	  zheng_two-dimensional_2023,
  author	= {Zheng, Xuchen and Chen, Siao and Li, Jinze and Wu, Han and
		  Zhang, Chao and Zhang, Danyan and Chen, Xi and Gao, Yang
		  and He, Feng and Hui, Lan and Liu, Huibiao and Jiu,
		  Tonggang and Wang, Ning and Li, Guoxing and Xu, Jialiang
		  and Xue, Yurui and Huang, Changshui and Chen, Chunying and
		  Guo, Yanbing and Lu, Tong-Bu and Wang, Dan and Mao, Lanqun
		  and Zhang, Jin and Zhang, Yue and Chi, Lifeng and Guo,
		  Wanlin and Bu, Xian-He and Zhang, Hongjie and Dai, Liming
		  and Zhao, Yuliang and Li, Yuliang},
  title		= {Two-{Dimensional} {Carbon} {Graphdiyne}: {Advances} in
		  {Fundamental} and {Application} {Research}},
  journal	= {ACS Nano},
  year		= {2023},
  volume	= {17},
  number	= {15},
  pages		= {14309--14346},
  month		= aug,
  issn		= {1936-0851},
  note		= {Publisher: American Chemical Society},
  doi		= {10.1021/acsnano.3c03849},
  shorttitle	= {Two-{Dimensional} {Carbon} {Graphdiyne}},
  urldate	= {2025-09-04}
}

@Article{	  research-prog-2d-mat-2025,
  author	= {Long, De-Bing and Zhao, Hongyun and Zhang, Jirong and Lei,
		  Yawen and Zhou, Zhishan and Zhang, Shuli and Yue, Ziying
		  and Peng, Shan and Wu, Xiaolin},
  title		= {Research progress in two-dimensional materials},
  journal	= {Journal of Materials Science},
  year		= {2025},
  month		= sep,
  issn		= {1573-4803},
  doi		= {10.1007/s10853-025-11392-6}
}

@Article{	  ares_recent_2022,
  author	= {Ares, Pablo and Novoselov, Kostya S.},
  title		= {Recent advances in graphene and other {2D} materials},
  journal	= {Nano Materials Science},
  year		= {2022},
  volume	= {4},
  number	= {1},
  pages		= {3--9},
  month		= mar,
  issn		= {2589-9651},
  doi		= {10.1016/j.nanoms.2021.05.002},
  series	= {Special issue on {Graphene} and {2D} {Alternative}
		  {Materials}}
}

@Article{	  mat-23-syn-chem-17,
  author	= {Mannix, Andrew J. and Kiraly, Brian and Hersam, Mark C.
		  and Guisinger, Nathan P.},
  title		= {Synthesis and chemistry of elemental 2D materials},
  journal	= {Nature Reviews Chemistry},
  year		= {2017},
  volume	= {1},
  number	= {2},
  pages		= {2397-3358},
  doi		= {10.1038/s41570-016-0014}
}

@Article{	  wang_helium_2025,
  author	= {Wang, Kun and Liao, Yi and Guo, Weijie and Liu, Xuming and
		  Pan, Changzhao},
  title		= {Helium isotope separation technology: {A} comprehensive
		  review and perspective},
  journal	= {Cryogenics},
  year		= {2025},
  volume	= {150},
  pages		= {104113},
  month		= sep,
  issn		= {0011-2275},
  doi		= {10.1016/j.cryogenics.2025.104113}
}

@Article{	  quant-effect-kin-sieving-05,
  author	= {Kumar, A. V. Anil and Bhatia, Suresh K.},
  title		= {Quantum Effect Induced Reverse Kinetic Molecular Sieving
		  in Microporous Materials},
  journal	= {Phys. Rev. Lett.},
  year		= {2005},
  volume	= {95},
  pages		= {245901},
  doi		= {10.1103/PhysRevLett.95.245901},
  issue		= {24}
}

@Article{	  rev-30years-istopic-sep-25,
  author	= {Sholl, David S.},
  title		= {Quantum Sieving for Isotopic Separations of Gases Using
		  Porous Materials─30 Years of Progress},
  journal	= {The Journal of Physical Chemistry C},
  year		= {2025},
  volume	= {129},
  number	= {24},
  pages		= {10793-10801},
  doi		= {10.1021/acs.jpcc.5c02981}
}

@Article{	  twisted-bilayer-jarillo-2018,
  author	= {Cao, Yuan and Fatemi, Valla and Demir, Ahmet and Fang,
		  Shiang and Tomarken, Spencer L. and Luo, Jason Y. and
		  Sanchez-Yamagishi, Javier D. and Watanabe, Kenji and
		  Taniguchi, Takashi and Kaxiras, Efthimios and Ashoori, Ray
		  C. and Jarillo-Herrero, Pablo},
  title		= {Correlated insulator behaviour at half-filling in
		  magic-angle graphene superlattices},
  journal	= {Nature},
  year		= {2018},
  pages		= {80-84},
  volume	= {556},
  doi		= {10.1038/nature26154}
}

@Article{	  gr2-multlayer-water-selec-25,
  author	= {Li, Jiaqiang and Zhou, Ke and Liu, Qing and Tian, Bo and
		  Liu, Xiaowei and Cao, Li and Cao, Haicheng and Li, Guanxing
		  and Zhang, Xixiang and Han, Yu and Lai, Zhiping},
  title		= {Synthesis of two-dimensional ordered graphdiyne membranes
		  for highly efficient and selective water transport},
  journal	= {Nature Water},
  year		= {2025},
  volume	= {3},
  pages		= {307–318},
  doi		= {10.1038/s44221-025-00397-9}
}

@Article{	  nanoscale-12-struct-bi-gr2-gr3,
  author	= {Zheng, Qiye and Luo, Guangfu and Liu, Qihang and Quhe,
		  Ruge and Zheng, Jiaxin and Tang, Kechao and Gao, Zhengxiang
		  and Nagase, Shigeru and Lu, Jing},
  title		= {Structural and electronic properties of bilayer and
		  trilayer graphdiyne},
  journal	= {Nanoscale},
  year		= {2012},
  volume	= {4},
  pages		= {3990-3996},
  doi		= {10.1039/C2NR12026G},
  issue		= {13}
}

@Article{	  lifetime-he-surf-td-94,
  author	= {Hern\'andez, Marta I. and Campos-Mart\'{\i}nez, Jos\'e and
		  Miret-Art\'es, S. and Coalson, Rob D.},
  title		= {Lifetimes of selective-adsorption resonances in
		  atom-surface elastic scattering},
  journal	= {Phys. Rev. B},
  year		= {1994},
  volume	= {49},
  pages		= {8300--8309},
  doi		= {10.1103/PhysRevB.49.8300},
  issue		= {12}
}

@Article{	  pgrogress-gr2-gas-separa-25,
  author	= {Siyuan Li and Yasong Zhao and Dan Wang},
  title		= {Progress in graphdiyne-based membrane for gas separation
		  and water purification},
  journal	= {ChemPhysMater},
  year		= {2025},
  volume	= {4},
  number	= {3},
  pages		= {234-250},
  issn		= {2772-5715},
  doi		= {10.1016/j.chphma.2025.02.007},
  url		= {https://www.sciencedirect.com/science/article/pii/S2772571525000208}
}

@Article{	  stacking-engi-2dmat-24,
  author	= {Fox, Carter and Mao, Yulu and Zhang, Xiang and Wang, Ying
		  and Xiao, Jun},
  title		= {Stacking Order Engineering of Two-Dimensional Materials
		  and Device Applications},
  journal	= {Chemical Reviews},
  year		= {2024},
  volume	= {124},
  number	= {4},
  pages		= {1862-1898},
  note		= {PMID: 38150266},
  doi		= {10.1021/acs.chemrev.3c00618},
  eprint	= {https://doi.org/10.1021/acs.chemrev.3c00618}
}

@Article{	  stacking-2dmat-21,
  author	= {Guo, Hao-Wei and Hu, Zhen and Liu, Zhi-Bo and Tian,
		  Jian-Guo},
  title		= {Stacking of 2D Materials},
  journal	= {Advanced Functional Materials},
  year		= {2021},
  volume	= {31},
  number	= {4},
  pages		= {2007810},
  doi		= {10.1002/adfm.202007810},
  eprint	= {https://advanced.onlinelibrary.wiley.com/doi/pdf/10.1002/adfm.202007810}
}

@Article{	  yang_intercalation_2024,
  author	= {Yang, Ruijie and Mei, Liang and Lin, Zhaoyang and Fan,
		  Yingying and Lim, Jongwoo and Guo, Jinghua and Liu, Yijin
		  and Shin, Hyeon Suk and Voiry, Damien and Lu, Qingye and
		  Li, Ju and Zeng, Zhiyuan},
  title		= {Intercalation in {2D} materials and in situ studies},
  journal	= {Nature Reviews Chemistry},
  year		= {2024},
  volume	= {8},
  number	= {6},
  pages		= {410--432},
  month		= jun,
  issn		= {2397-3358},
  note		= {Publisher: Nature Publishing Group},
  doi		= {10.1038/s41570-024-00605-2},
  language	= {en},
  url		= {https://www.nature.com/articles/s41570-024-00605-2},
  urldate	= {2025-10-29}
}

@Article{	  doping-2dmat-17,
  author	= {Feng, Simin and Lin, Zhong and Gan, Xin and Lv, Ruitao and
		  Terrones, Mauricio},
  title		= {Doping two-dimensional materials: ultra-sensitive
		  sensors{,} band gap tuning and ferromagnetic monolayers},
  journal	= {Nanoscale Horiz.},
  year		= {2017},
  volume	= {2},
  pages		= {72-80},
  doi		= {10.1039/C6NH00192K},
  issue		= {2}
}

@Article{	  recent-advan-2dmat-batteries-24,
  author	= {Yinghui Xue and Tianjie Xu and Chenyang Wang and Lei Fu},
  title		= {Recent advances of two-dimensional materials-based
		  heterostructures for rechargeable batteries},
  journal	= {iScience},
  year		= {2024},
  volume	= {27},
  number	= {8},
  pages		= {110392},
  issn		= {2589-0042},
  doi		= {10.1016/j.isci.2024.110392},
  url		= {https://www.sciencedirect.com/science/article/pii/S2589004224016171}
}

@Article{	  gr2-baterias-max-24,
  author	= {Bartolomei, Massimiliano and Giorgi, Giacomo},
  title		= {Sodium into γ-Graphyne Multilayers: An Intercalation
		  Compound for Anodes in Metal-Ion Batteries},
  journal	= {ACS Materials Letters},
  year		= {2024},
  volume	= {6},
  number	= {10},
  pages		= {4682-4689},
  doi		= {10.1021/acsmaterialslett.4c01119}
}

@Article{	  he-isotopic-gr3-wp3d-21,
  author	= {Hernández, Marta I. and Bartolomei, Massimiliano and
		  Campos-Martínez, José},
  title		= {Helium Isotopes Quantum Sieving through Graphtriyne
		  Membranes},
  journal	= {Nanomaterials},
  year		= {2021},
  volume	= {11},
  number	= {1},
  pages		= {73},
  issn		= {2079-4991},
  doi		= {10.3390/nano11010073}
}

@Article{	  ilj,
  author	= {F. Pirani and S. Brizi and L. Roncaratti and P.
		  Casavecchia and D. Cappelletti and F. Vecchiocattivi},
  title		= {Beyond the $\rm{Lennard}$-$\rm{Jones}$ Model: A Simple and
		  Accurate Potential Function Probed by High Resolution
		  Scattering Data Useful for Molecular Dynamics Simulations},
  journal	= {Phys. Chem. Chem. Phys},
  year		= {2008},
  volume	= {10},
  pages		= {5489-5503}
}

@Article{	  miller:74,
  author	= {W. H. Miller},
  title		= {Quantum Mechanical Transition State Theory and a New
		  Semiclassical Model for Reaction Rate Constants},
  journal	= {J. Chem. Phys.},
  year		= {1974},
  volume	= {61},
  pages		= {1823-1834}
}

@Article{	  reider-ijhe-2024,
  author	= {Reider, Anna Maria and Kollotzek, Siegfried and Scheier,
		  Paul and Calvo, Florent and Yurtsever, Ersin and Pirani,
		  Fernando and Bartolomei, Massimiliano and Hern{\'a}ndez,
		  Marta I and Gonz{\'a}lez-Lezana, Tom{\'a}s and
		  Campos-Mart{\'\i}nez, Jos{\'e}},
  title		= {Experimental and theoretical assessment of the enhanced
		  hydrogen adsorption on polycyclic aromatic hydrocarbons
		  upon decoration with alkali metals},
  journal	= {Int. J. Hydrog. Energy},
  year		= {2024},
  volume	= {58},
  pages		= {525--535},
  doi		= {10.1016/j.ijhydene.2024.01.244}
}

@Article{	  gr2-rev-fundamental-and-applied,
  author	= {Zheng, Xuchen and Chen, Siao and Li, Jinze and Wu, Han and
		  Zhang, Chao and Zhang, Danyan and Chen, Xi and Gao, Yang
		  and He, Feng and Hui, Lan and Liu, Huibiao and Jiu,
		  Tonggang and Wang, Ning and Li, Guoxing and Xu, Jialiang
		  and Xue, Yurui and Huang, Changshui and Chen, Chunying and
		  Guo, Yanbing and Lu, Tong-Bu and Wang, Dan and Mao, Lanqun
		  and Zhang, Jin and Zhang, Yue and Chi, Lifeng and Guo,
		  Wanlin and Bu, Xian-He and Zhang, Hongjie and Dai, Liming
		  and Zhao, Yuliang and Li, Yuliang},
  title		= {Two-Dimensional Carbon Graphdiyne: Advances in Fundamental
		  and Application Research},
  journal	= {ACS Nano},
  year		= {2023},
  volume	= {17},
  number	= {15},
  pages		= {14309-14346},
  note		= {PMID: 37471703},
  doi		= {10.1021/acsnano.3c03849},
  eprint	= {https://doi.org/10.1021/acsnano.3c03849},
  url		= {https://doi.org/10.1021/acsnano.3c03849 }
}

@Article{	  noauthor_doping_2021,
  title		= {Doping in {2D}},
  journal	= {Nature Electronics},
  year		= {2021},
  volume	= {4},
  number	= {10},
  pages		= {699--699},
  month		= oct,
  issn		= {2520-1131},
  note		= {Publisher: Nature Publishing Group},
  doi		= {10.1038/s41928-021-00668-9}
}

@Article{	  heteroatomdop-2dma-20,
  author	= {Haoyue Zhu and Xin Gan and Amber McCreary and Ruitao Lv
		  and Zhong Lin and Mauricio Terrones},
  title		= {Heteroatom doping of two-dimensional materials: From
		  graphene to chalcogenides},
  journal	= {Nano Today},
  year		= {2020},
  volume	= {30},
  pages		= {100829},
  issn		= {1748-0132},
  doi		= {https://doi.org/10.1016/j.nantod.2019.100829},
  url		= {https://www.sciencedirect.com/science/article/pii/S1748013219302294}
}

@Article{	  beenakker-prl-94,
  title		= {Molecular transport in the nanometer regime},
  author	= {Beenakker, J. J. M. and Borman, V. D. and Krylov, S. Yu.},
  journal	= {Phys. Rev. Lett.},
  volume	= {72},
  issue		= {4},
  pages		= {514--517},
  numpages	= {0},
  year		= {1994},
  month		= {Jan},
  publisher	= {American Physical Society},
  doi		= {10.1103/PhysRevLett.72.514},
  url		= {https://link.aps.org/doi/10.1103/PhysRevLett.72.514}
}

@Article{	  beenakker-cpl-95,
  title		= {Molecular transport in subnanometer pores: zero-point
		  energy, reduced dimensionality and quantum sieving},
  journal	= {Chemical Physics Letters},
  volume	= {232},
  number	= {4},
  pages		= {379-382},
  year		= {1995},
  issn		= {0009-2614},
  doi		= {https://doi.org/10.1016/0009-2614(94)01372-3},
  url		= {https://www.sciencedirect.com/science/article/pii/0009261494013723},
  author	= {J.J.M. Beenakker and V.D. Borman and S.Yu. Krylov}
}

@Article{	  freeman-60-quantum,
  author	= {Freeman, Mark P.},
  title		= {THE QUANTUM MECHANICAL CORRECTION FOR THE HIGH TEMPERATURE
		  VAN DER WAALS INTERACTION OF LIGHT GASES AND SURFACES. A
		  NEW METHOD OF DETERMINING SURFACE AREA},
  journal	= {The Journal of Physical Chemistry},
  volume	= {64},
  number	= {1},
  pages		= {32-37},
  year		= {1960},
  doi		= {10.1021/j100830a008},
  url		= { https://doi.org/10.1021/j100830a008}
}

@Article{	  quantum-siev-12-rsc,
  author	= "Cai, Jinjun and Xing, Yanlong and Zhao, Xuebo",
  title		= "Quantum sieving: feasibility and challenges for the
		  separation of hydrogen isotopes in nanoporous materials",
  journal	= "RSC Adv.",
  year		= "2012",
  volume	= "2",
  issue		= "23",
  pages		= "8579-8586",
  publisher	= "The Royal Society of Chemistry",
  doi		= "10.1039/C2RA01284G",
  url		= "http://dx.doi.org/10.1039/C2RA01284G"
}

@Article{	  synthe-graphy-22-classifi,
  author	= {Desyatkin, Victor G. and Martin, William B. and Aliev, Ali
		  E. and Chapman, Nathaniel E. and Fonseca, Alexandre F. and
		  Galvão, Douglas S. and Miller, Ericka Roy and Stone, Kevin
		  H. and Wang, Zhong and Zakhidov, Dante and Limpoco, F. Ted
		  and Almahdali, Sarah R. and Parker, Shane M. and Baughman,
		  Ray H. and Rodionov, Valentin O.},
  title		= {Scalable Synthesis and Characterization of Multilayer
		  γ-Graphyne, New Carbon Crystals with a Small Direct Band
		  Gap},
  journal	= {Journal of the American Chemical Society},
  volume	= {144},
  number	= {39},
  pages		= {17999-18008},
  year		= {2022},
  doi		= {10.1021/jacs.2c06583},
  url		= {https://doi.org/10.1021/jacs.2c06583}
}

@Article{	  mp2c,
  author	= {M. Piton\'ak and A. Hesselmann},
  title		= {Accurate Intermolecular Interaction Energies from a
		  Combination of MP2 and TDDFT Response Theory},
  journal	= {J. Chem. Theory Comput.},
  pages		= {168-178},
  year		= 2010,
  volume	= 6
}

@Article{	  water-gr2-permea-our-14,
  author	= {Bartolomei, Massimiliano and Carmona-Novillo, Estela and
		  Hernández, Marta I. and Campos-Martínez, Jos{\'e} and
		  Pirani, Fernando and Giorgi, Giacomo and Yamashita, Koichi},
  title		= {Penetration Barrier of Water through Graphynes’ Pores:
		  First-Principles Predictions and Force Field Optimization},
  journal	= {The Journal of Physical Chemistry Letters},
  volume	= {5},
  number	= {4},
  pages		= {751-755},
  year		= {2014},
  doi		= {10.1021/jz4026563},
  url		= { https://doi.org/10.1021/jz4026563}
}




\newpage

\appendix
\section{Supplementary Information}

\subsection{Bilayer Graphdiyne}

\begin{table}[h]
  \centering
\begin{tabular}{|c|rrr|}
  \hline
    Atom &     $x$ (\AA)  &    $y$ (\AA) &  $z$ (\AA)     \\
    \hline
\hline
  C   &-2.728299 &  3.294551  & 0.000000 \\
  C   &-1.489016 &  4.010052  & 0.000000 \\
  C   &-3.967582 &  4.010052  & 0.000000 \\
  C   &-7.399694 &  2.028521  & 0.000000 \\
  C   &-6.241801 &  2.697031  & 0.000000 \\
  C   &-5.175702 &  3.312544  & 0.000000 \\
  C   &-1.489016 & -4.010052  & 0.000000 \\
  C   &-2.728299 & -3.294551  & 0.000000 \\
  C   &-3.967582 & -4.010052  & 0.000000 \\
  C   &-7.399694 & -2.028521  & 0.000000 \\
  C   &-6.241801 & -2.697031  & 0.000000 \\
  C   &-5.175702 & -3.312544  & 0.000000 \\
  C   & 5.456598 & -1.431001  & 0.000000 \\
  C   & 6.695881 & -0.7155005 & 0.000000 \\
  C   & 6.695881 &  0.7155005 & 0.000000 \\
  C   & 5.456598 &  1.431001  & 0.000000 \\
  C   & 4.217315 &  0.7155005 & 0.000000 \\
  C   & 4.217315 & -0.7155005 & 0.000000 \\
  C   &-2.728299 & -0.6685095 & 0.000000 \\
  C   &-2.728299 &  0.6685095 & 0.000000 \\
  C   &-0.2808957& -3.312543  & 0.000000 \\
  C   & 0.7852032& -2.697031  & 0.000000 \\
  C   &-0.2808957&  3.312543  & 0.000000 \\
  C   & 0.7852032&  2.697031  & 0.000000 \\
  C   & 1.943096 & -2.028522  & 0.000000 \\
  C   & 3.009195 & -1.413009  & 0.000000 \\
  C   & 1.943096 &  2.028522  & 0.000000 \\
  C   & 3.009195 &  1.413009  & 0.000000 \\
  C   &-2.728299 & -1.899534  & 0.000000 \\
  C   &-2.728299 &  1.899534  & 0.000000 \\
  C   & 5.456598 &  4.057043  & 0.000000 \\
  C   & 7.904001 &  1.413009  & 0.000000 \\
  C   & 5.456598 &  2.826018  & 0.000000 \\
  C   & 5.456598 & -4.057043  & 0.000000 \\
  C   & 7.904001 & -1.413009  & 0.000000 \\
  C   & 5.456598 & -2.826018  & 0.000000 \\
\hline\hline    
\end{tabular}
    \caption{  Cartesian coordinates unit cell of single layer Graphdiyne.}
    \label{geom-gr2}
    \end{table}


\begin{table}[h]
       \centering
\begin{tabular}{|c|rrr|}
 \hline
 Bilayer      &     $x$ (\AA)  &    $y$ (\AA) &  $z$ (\AA)  \\  [-1mm]
    \hline
 1st.layer    &          0.    &   0.         &  0.         \\  [-1mm]
    \hline
    \hline
    $AA$     &          0.    &   0.         &  3.65   \\  [-1mm]
    \hline
    $AA\; -1$    &          0.    &   0.         &  2.5   \\  [-1mm]
     $AA\; -2$    &          0.    &   0.         &  3.5   \\  [-1mm]
 $AB$         &          1.45  &   0.        &  3.45  \\  [-1mm]
   \hline\hline
  \end{tabular}
    \caption[]{ Cartesian coordinates of the center of pore for bilayer
    graphdiyne systems. First column indicates the type of stacking of the two
    layers. The larger separation in  $AA$ stacking corresponds to a distance
    of $3.65$ \AA $^{S1}$, more unstable than other (such
    as $AB$) stackings due to electron cloud repulsion.
    Two additional $AA$ stackings corresponding
    to interlayer separations of $2.5$ \AA $\,$ and $3.5$ \AA, have also
    been studied.
    The values taken in this work for $AB$ are similar
    to those of Ref.[S2] for graphtriyne.}
      \label{geom-bilayer}
\end{table}

\begin{figure*}
  \includegraphics[width=8.3cm]{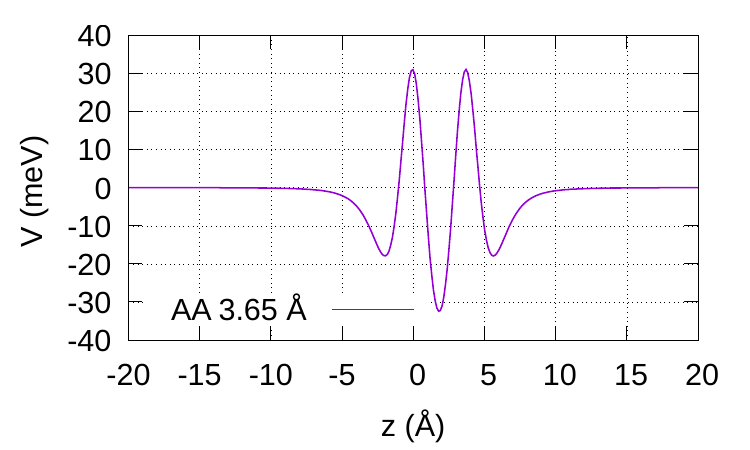}%
  \caption{Interaction Potential in a perpendicular approach along
    $Z$ axis, for a bilayer $AA$ nanostructure separated by a distance
    of 3.65 $\text{\AA}$. }
      \label{pot-per-AA}
\end{figure*}

\begin{figure*}
  \includegraphics[width=8.3cm]{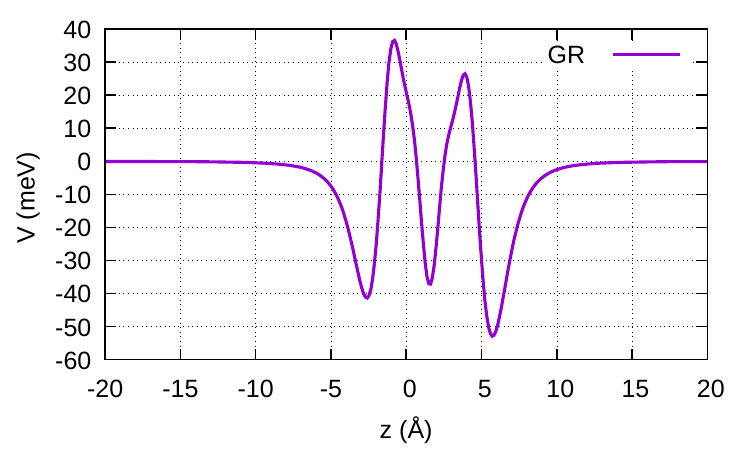}%
  \caption{Interaction Potential for the nanostructure of $AB$ stacking
    in Table \ref{geom-bilayer} in the direction of minimal
    barrier ($\theta\approx \;27^0 $, incident angle ).}
      \label{pot-per-il-2.5}
\end{figure*}

\begin{figure*}
  \includegraphics[width=8.3cm]{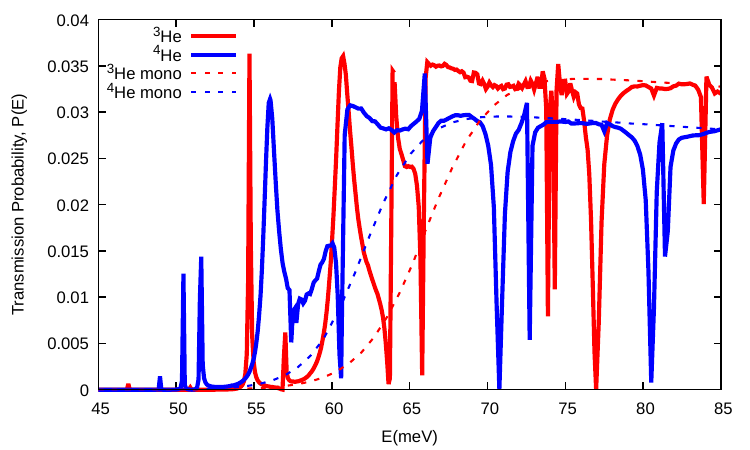}
  \caption{Probabilities for  ($^3He/^4He$) transmission through
    bilayer graphdiyne for $AA$ stacking and $3.65\,\text{\AA}$ interlayer
    separation.  The narrower energy window allow us to easily watch
    the maxima and minima corresponding in transmission probabilities
    corresponding to Fig.3 in the main text.  Dashed lines for the
    monolayer case are included.}
  \label{figs5}
\end{figure*}

\begin{figure*}
  \includegraphics[width=8.3cm]{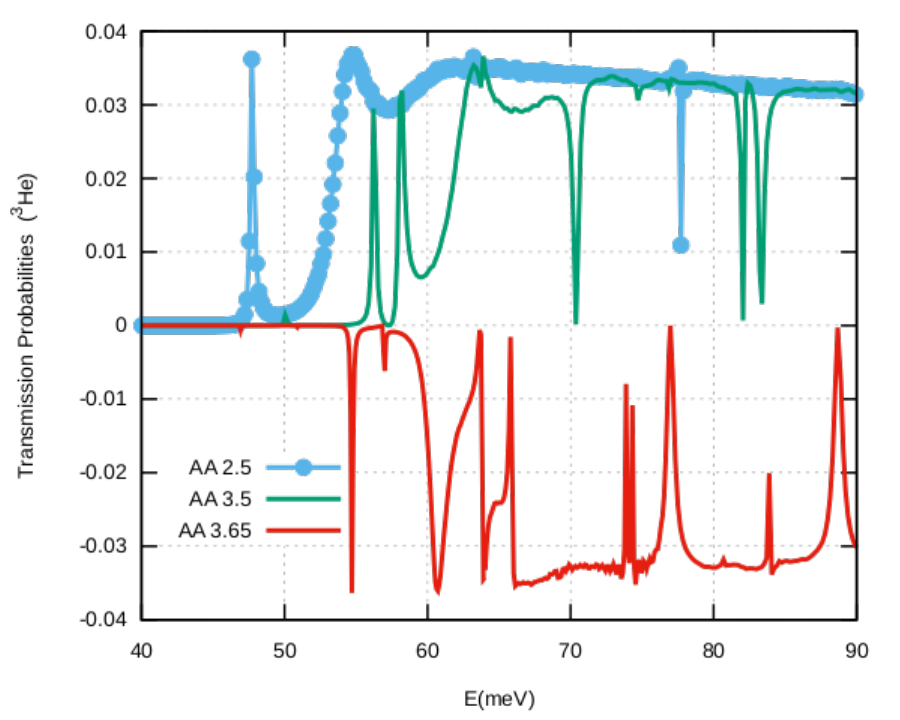}
  \caption{Probabilities of He ($^3He$) transmission through bilayer graphdiyne
    for $AA$ stacking at three different values of interlayer separation
    and for a narrower range of incident energies.
    For the $3.65\,\text{\AA}$ interlayer separation, we have plotted the
    probabilities with negative value to better watch the change in
    the positions of the spikes.}
  \label{figs6}
\end{figure*}

\subsection{Interaction Potential }
The total interaction between $He$ atoms and the Graphdiyne (GDY) layers is
described as a sum of $He-C$,  $ V = \sum_i V^{i} $, summed over all
carbon atoms in the membrane, and where
each contribution, $i$, is described by the so-called Improved
Lennard-Jones  $^{S3}$  (ILJ) 

\begin{equation}
V^{i}(R)=
\varepsilon\left[\frac{6}{n(x)-6}\left (
\frac{1}{x} \right)^{n(x)} -
\frac{n(x)}{n(x)-6}\left(
\frac{1}{x}\right)^6 \right]
\label{eq-ilj}
\end{equation}

\noindent
where $x$ is a reduced pair distance given by $x = \frac{R}{R_{m}}$,
and where $\varepsilon$ and $R_{m}$ represent the well depth and equilibrium
distance respectively.  In Eq.(\ref{eq-ilj}) $n(x)$ follows a
expression $n(x)= \beta + 4.0 \; x^2$ where $\beta$ is a parameter
defining the shape and stiffness of the potential.

The parameters have been already given in Ref.[S4],
and used for a single layer calculation in Ref.[S5].  Here
the sum has to be extended to the two GDY layers.

\vspace{0.2cm}

\noindent
\subsection{Three-dimensional time-dependent wave packet propagation:
description and computational details} 

The quantum mechanical procedure applied to the transmission of atoms
through 2D layers, was originally described in Ref.[S5],
therefore here we will give a short summary of such a procedure
for the sake of clarity.

The total Hamiltonian for the the two isotopes, $^3He,\; ^4He$ of helium
interacting with  a GDY membrane is written as

\begin{equation}
  H = \frac{-\hbar^2}{2m} \left[\, \frac{\partial^2}{\partial x^2} +
    \frac{\partial^2}{\partial y^2} +
    \frac{\partial^2}{\partial z^2} \right]\, + V(x,y,z),
\end{equation}

\noindent
where $V$ is the interaction potential just described, and $m$ the mass corresponding to
each isotope. We use periodic conditions, as described by Yinnon and
Kosloff$^{S6}$, limiting the coordinates range  $x$ and $y$
to the length of the unit-cell.   For the coordinate $z$, the grid should
be large enough to contain all desired energies of the wave packet and to
describe the approach of the wave packet and the reflection and
transmission.

The initial wave packet is taken as,

\begin{equation}
  \psi (x,y,z,t=0) = G(z;z_0,k_{z0},alpha) \,
  \frac{ \exp \left[i {\bf K \cdot R}\right] } {(\Delta_x \Delta_y)^{1/2}}, 
  \label{initialwp}
\end{equation}

\noindent
in this expression the exponential function is a plane wave with a wave
vector ${\bf K}$, and position vector  ${\bf R}$ in the plane $XY$,
$\Delta_x$ and $\Delta_y$ are the lengths of the unit cell
($\Delta_x= \sqrt{3}\Delta_y$, $\Delta_y$= 9.45 \AA, see Table \ref{geom-gr2}).
The values of ${\bf K}$ must be chosen such that the plane wave is
periodic with the lattice$^{S6}$, a condition achieved when 
\begin{equation}
   {\bf K} = 2 \pi \left( \frac{ l_x}{\Delta_x}, \frac{ l_y}{\Delta_y} \right)
\end{equation}

\noindent
and $l_x,l_y$ are integer numbers.  In our work for $AA$ stacking
in Table \ref{geom-bilayer}, ${\bf K}=0$, that implies a perpendicular
approach to the 2D material. For the $AB$ stacking case, these integers
are conveniently chosen to get an adequate incident angle for the
initial wave packet.

The initial wave packet along the $Z$ direction is a Gaussian,

\begin{equation}
  G(z;z_0,k_{z0},\alpha) = \left( \frac{2
    \operatorname{Im}(\alpha) }{\pi \hbar} \right)^{1/4}
 \exp \left[i \alpha (z-z_0)^2/\hbar + i k_{z0}(z-z_0) \right]  
\label{eq2}
\end{equation}

\noindent
where $z_o$ and $k_{z0}$ are the central values of the wave
packet in the position and momentum spaces, respectively, while $\alpha$ is a
pure imaginary number that determines the width of the Gaussian function. Parameters
used for the Gaussian wavepacket are given in Table \ref{GWP-parameters}.

\begin{table}[h]
\centering
\begin{tabular}{|c|c|c|}
\hline
\multicolumn{3}{|c|}{Gaussian} \\
\hline
$Im(\alpha)\,\frac{[\hbar]}{[L^2]}$ & $z_0$(\AA) & $k_{z_0}$(\AA$^{-1}$)  \\
\hline
  0.3882     & 35.   & [6.0 - 15.0]  \\
\hline
\end{tabular}
\caption{Parameters of the Gaussian wave packet propagation. $\alpha$ is
  a parameter in atomic units that to be dimensionally congruent is
 $[M].[T]^{-1}$.}

\label{GWP-parameters}
\end{table}

The grid in $Z$ is much larger than in the direction of the plane $XY$ (the
GDY layers), to allow for regions where the wave packet is considered
to be in the asymptotic region.  The
wave packet is initially placed in this region, and the it is propagated with
the Split-Operator method$^{S7,s8}$.
The propagation needs to
be maintained for a long time, and absorbing boundary conditions need
to be applied$^{S9,S10}$.

\begin{table}[h]
\centering
\begin{tabular}{|c|c|c| c|}
\hline
                    & $x$               &      $y$       & $z$         \\
\hline
Box (\AA)           &  (-8.185, 8.185 ) & (-4.725,4.725) & ( -30.0-35.0, 55.0-60.0)\\
\hline
Number of points    & 128               & 128            & 1024   \\
\hline
\end{tabular}
\caption{Grid sizes and number of points for the representation of the
  wave packet.}
  \label{Grid}
\end{table}

\small

\begin{table}[h]
\centering
\begin{tabular}{|c|c|c|c|c|c|c|}
\hline
\multicolumn{2}{|c|}{Time}          & \multicolumn{4}{|c|}{Wave packet} & Flux    \\
\multicolumn{2}{|c|}{propagation}   & \multicolumn{4}{|c|}{splitting}   & surface \\
\hline
$\delta t$($fs$) & $t_{final}$($ps$) & $\Delta t$($fs$) & $z_+$(\AA) & $z_-$(\AA) & $\beta$(\AA$^{-2}$)
& $z_f$(\AA)  \\
\hline
   0.083   & [30.0-40.0] & 0.326 & [45-50] &[-25.0,-20.0] & [0.01-0.05]  & [-3.5, -4.5] \\
   \hline
\end{tabular}
\caption{Parameters of the wave packet propagation, [: - :]
 indicates the range of values taken in different calculations
   (definitions specified in the text).}
\label{WP-parameters}
\end{table}

\normalsize

 Absorbing boundary conditions$^{S9,S10}$
were applied for the non-periodic coordinate $(z)$  in the regions
defined by  $z< z_{-}$ and $z> z_+$ and with a time period $\Delta t$, the
wave packet is split into interaction and product wave packets 
using the damping function $\exp{[-\beta(z-z_{\pm})^2]}$.
After splitting, propagation is resumed using the interaction portion of the
wave packet.

In order to obtain the transmission probability it is necessary
to compute the stationary wave function and its derivative at the flux surface $z=z_f$.
This task is performed by accumulating the integrand of the time-energy Fourier
transform along the propagation time$^{S11}$. 
Computations run until most of the wave packet gets out of the
interaction region, remaining only a small portion in this
region ($\approx 1 $\%), for calculations of probabilities shown in
the main text.
Total propagation times $t_{final}$ range between
the values indicated in Table \ref{WP-parameters}, the shortest (largest)
times corresponding to the calculations at largest (smaller) energies
in the initial wavepacket.

Values of all parameters of the calculations mentioned above are gathered in
Table \ref{WP-parameters}. Several convergence checks were carried out by varying
these values as well as the number of grid points and the $z$ box size.

REFERENCES:

(S1) Zheng, Q.; Luo, G.; Liu, Q.; Quhe, R.; Zheng, J.; Tang, K.; Gao, Z.; Nagase, S.;
Lu, J. Structural and Electronic Properties of Bilayer and Trilayer Graphdiyne.
Nanoscale 2012, 4, 3990–3996.

(S2) Bartolomei, M.; Giorgi, G. a Novel Nanoporous Graphite Based on Graphynes:
First-Principles Structure and Carbon Dioxide Preferential Physisorption. ACS
Appl. Mater. Interfaces 2016, 8, 27996–28003.

(S3) Pirani, F.; Brizi, S.; Roncaratti, L.; Casavecchia, P.; Cappelletti, D.;
Vecchiocattivi, F. Beyond the Lennard-Jones Model: A Simple and Accurate Potential Func-
tion Probed by High Resolution Scattering Data Useful for Molecular Dynamics
Simulations. Phys. Chem. Chem. Phys 2008, 10, 5489–5503.

(S4) Bartolomei, M.; Carmona-Novillo, E.; Hernández, M. I.;
Campos-Mart{\'i}nez, J.; Pirani, F.; Giorgi, G. Graphdiyne pores: Ad hoc
openings for helium separation applications. J. Phys. Chem. C 2014, 118,
29966–29972.

(S5) Gijón, A.; Campos-Mart{\'i}nez, J.; Hernández, M. Wave Packet
Calculations of the Quantum Transport of Atoms Through Nanoporous Membranes.
J. Phys. Chem. C 2017, 121, 19751–19757.

(S6) Yinnon, A. T.; Kosloff, R. A Quantum-Mechanical Time-Dependent
Simulation of the Scattering from a Stepped Surface. Chem. Phys. Lett. 1983,
102, 216–223.

(S7) Feit, M. D.; Jr., J. A. F.; Steiger, A. Solution of the Schr\"odinger
Equation by a Spectral Method. J. Comput. Phys. 1982, 47, 412–433.

(S8) Kosloff, D.; Kosloff, R. A Fourier Method Solution for the Time Dependent
Schr\"odinger Equation as a Tool in Molecular Dynamics. J. Comp. Phys. 1983,
52, 35–53.

(S9) R. Heather and Horia Metiu., J. Chem. Phys. 1987, 86, 55009.

(S10) Pernot, P.; Lester, W. A. Multidimensional Wave-Packet Analysis: Splitting Method
for Time-Resolved Property Determination. Int. J. Quantum Chem. 1991, 40, 577–
588.

(S11) di Domenico, D.; Hernández, M. I.; Campos-Mart{\'i}nez, J. A Time-Dependent Wave
Packet Approach for Reaction and Dissociation in H2+H2 . Chem. Phys. Lett. 2001,
342, 177–184.

\end{document}